\documentclass[aps,prc,twocolumn,showpacs,amsmath,amssymb,nofootinbib]{revtex4-2}
\usepackage{graphicx}  
\usepackage{color}
\usepackage{times}
\usepackage{inputenc}
\usepackage{float}
\usepackage{bm}
\usepackage{ulem}
\usepackage{multirow}
\usepackage{float}
\usepackage{url}
\usepackage{natbib}
\usepackage{tabularx}

\usepackage{mathrsfs}
\usepackage{physics}
\usepackage{comment}
\usepackage[table,xcdraw]{xcolor}
\usepackage{booktabs,makecell}
\usepackage{comment}
\usepackage{amsmath}
\usepackage{pifont}
\usepackage{appendix}
\usepackage[colorlinks=true,citecolor=blue,urlcolor=blue,linkcolor=blue]{hyperref}
\begin{document}
\title{Universality in Space–Time $\omega$ modes of Quarkyonic Stars }
\author{D. Dey$^{1,2}$}
\email{debabrat.d@iopb.res.in }
\author{Jeet Amrit Pattnaik$^{3,4}$}
\email{jeetamritboudh@gmail.com }
\author{R. N. Panda$^{4}$}
\author{S. K. Patra$^{4}$}

\affiliation{\it $^{1}$Institute of Physics, Sachivalaya Marg, Bhubaneswar-751005, India}
\affiliation{\it $^{2}$Homi Bhabha National Institute, Training School Complex, 
Anushakti Nagar, Mumbai 400094, India}
\affiliation{\it $^{3}$Department of Physics, Indira Gandhi Institute of Technology,
Sarang, Dhenkanal, Odisha-759146, India}
\affiliation{\it $^{4}$Department of Physics, Siksha $'O'$ Anusandhan, Deemed to be University, Bhubaneswar -751030, India}
\date{\today}
\begin{abstract}
The gravitational wave $\omega$ mode spectrum presents a unique window into the dense interior of neutron stars, probing physics inaccessible to electromagnetic observations. This work investigates the  $\omega$ modes of compact stars composed of quarkyonic matter. The quarkyonic model, which describes a cross-over transition between nucleonic and quark matter treated as quasi-particles, is formulated within the Relativistic Mean-Field (RMF) theory using the G3 and IOPB-I parameterizations. This core is surrounded by a mantle of hadronic matter, creating a multicomponent stellar interior. The overall Equation of State (EOS) is governed by two key parameters: the transition density ($n_t$), the QCD confinement scale ($\Lambda_{\rm cs}$), which are varied to construct models consistent with current astrophysical constraints on mass and radius. We compute the complex eigenfrequencies (damped oscillations) of the fundamental and first excited  $\omega$ modes using the phase-amplitude method within a full general relativistic framework. Our simulations reveal that the admixed quarkyonic structure produces a unique $\omega$ mode signature, distinctly different from pure hadronic or hybrid stars. The spectrum exhibits a strong, degenerate dependence on the EOS, where the stiffening effect of the quarkyonic matter influences oscillation frequencies and damping times in a characteristic manner. We also demonstrate that $\omega$ mode frequencies for quarkyonic stars follow approximate universal relations, largely independent of the EOS. 
\end{abstract}
\maketitle
\section{Introduction}

The last decade has established neutron stars (NSs) as precision laboratories for dense-matter physics through a convergence of radio, X-ray, and gravitational-wave (GW) observations. In the radio band, the discovery of $\gtrsim 2,M_\odot$ pulsars has imposed a robust lower bound on the maximum mass supported by the dense-matter equation of state (EOS) \cite{Demorest_2010,Antoniadis_2013,Cromartie_2020,Romani_2022}. In the X-ray band, pulse-profile modeling with NICER has provided simultaneous constraints on mass and radius for multiple sources, directly informing the stiffness of the EOS around a few times nuclear saturation density \cite{Riley_2019,Miller_2019,Miller_2021}. In the GW band, the binary-neutron-star inspiral event GW170817 and subsequent analyses have constrained the tidal deformability of canonical-mass stars and have significantly reduced the viable range of EOS models \cite{GW170817_binary_props,PhysRevLett.121.091102,PhysRevLett.121.161101,Capano2020}. Together, these developments demonstrate that the EOS must be stiff enough to support heavy pulsars while remaining consistent with radius and tidal constraints inferred from multimessenger data \cite{Capano2020,Annala_2018}.

From an astrophysical standpoint, neutron stars are formed in the aftermath of core-collapse supernovae, when the degenerate stellar core collapses into a self-gravitating remnant supported primarily by neutron degeneracy pressure and strong-interaction repulsion \cite{Burrows2000,Lattimer_2004}. Their typical masses are of order $\sim 2 M_\odot$ and their radii are of order $\sim 10$--$14,\mathrm{km}$ \cite{Lattimer_2004,Lattimer_2007}, implying central densities that can reach several times the nuclear saturation density. In this regime, the interplay between general relativity and nuclear microphysics becomes essential: relativistic gravity determines the global structure, while the underlying strong-interaction physics controls the internal pressure support, composition, and transport properties \cite{Lattimer_2004}.

A central quantity in any theoretical description of neutron stars is the EOS, i.e., the relation between pressure and energy density, $p=p(\epsilon)$, for cold catalyzed matter in beta equilibrium \cite{Lattimer_2004}. Once an EOS is specified, the equilibrium stellar sequence (mass--radius relation) and the maximum stable mass follow from the Tolman--Oppenheimer--Volkoff (TOV) equations \cite{PhysRev.55.364,PhysRev.55.374}. Conversely, the observational program outlined above provides complementary integral constraints on the EOS; however, because the relevant densities are beyond those probed in laboratories, the EOS at supranuclear density remains uncertain \cite{Lattimer_2007}. This uncertainty propagates to key questions such as the density dependence of the sound speed, the onset of new degrees of freedom, and the nature of the hadron--quark transition in cold neutron-star matter \cite{Bedaque_2015,Masuda2013}.

Recent GW observations have sharpened the long-standing question of the so-called ``lower mass gap" between the heaviest neutron stars and the lightest black holes \cite{Abbott_2020}. The highly asymmetric merger GW190814 revealed a system containing a $\sim 23 M_\odot$ primary black hole and a compact secondary with mass $\sim 2.6 M_\odot$ \cite{Abbott_2020}. The nature of the secondary (an unusually massive neutron star versus a low-mass black hole) remains debated and depends sensitively on the unknown maximum neutron-star mass and on assumptions about spin and formation channels \cite{Most_2020,Huang_2020}. More recently, the event GW230529 \cite{jeetGW230529} was reported as a merger consistent with a neutron star and a compact companion with a mass in the approximate range $\sim 2.5$--$4.5 M_\odot$ \cite{Abac_2024}, again overlapping the putative mass-gap regime. These events provide a unique window into the microphysics of dense matter inside compact objects. In particular, they suggest the possible occurrence of phase transitions—such as the conversion of nuclear matter into deconfined quark matter—deep within neutron-star cores. Such microphysical effects could raise the maximum neutron-star mass and help explain the existence of unusually heavy neutron-star candidates that fall in the so-called lower mass gap \cite{Fattoyev_2020,Tan_2020}. 

From the theoretical side, due to the ultra-dense matter and complex structure of NS, many EOS models have been proposed, ranging from purely nucleonic descriptions to scenarios that incorporate additional degrees of freedom such as hyperons, meson condensates, or deconfined quark matter \cite{Lattimer_2007,schaffner-bielich_2020,Baym_2018}. A key question is not only whether quark degrees of freedom appear in NS interiors, but also how they emerge: through a first-order phase transition with a sharp interface and possible mixed phases, or through a smoother crossover \cite{Masuda2013,NKGfp_1992,eos_mixed_phase_2011,eos_mixed_phase_2016,eos_phasetransition_2024,eos_QCD_constraints_2022,eos_QCD_constraints_2023,Fujimoto_merger_2025}. In some first-order transition scenarios, the microphysics of the hadron--quark interface (including the conversion dynamics) can qualitatively affect stellar stability and the structure of equilibrium sequences \cite{NKGfp_1992}. In particular, it has been discussed that slow conversion at the phase-splitting surface can lead to extended branches of stable hybrid configurations, with important consequences for proposed universal relations and for the interpretation of global observables \cite{Orsaria_2014}.

In the present work, instead of a sharp boundary between hadronic and quark phases (first-order phase transition) as proposed in the original model \cite{McLerran_2019}, we focus on quarkyonic matter as a physically motivated crossover-type scenario between hadronic and quark matter at high density \cite{McLerran_2019,Koch_QM_2024}. It is worth noting that the present method differs slightly from the procedure proposed in \cite{patra07a, patra07b}, which is based on the Maxwell or Gibbs criterion. In quarkyonic models, confinement persists near the Fermi surface while quarks populate low-momentum states in the deep core; nucleons remain effective degrees of freedom in a shell near the Fermi surface, and quarks occupy the interior of momentum space \cite{McLerran_2019,tinaki_ur}. This construction can generate characteristic stiffening behavior at intermediate densities and nontrivial changes in the sound-speed profile while remaining consistent with expected asymptotic behavior at very high density \cite{McLerran_2019}. To connect nuclear microphysics to macroscopic observables in a systematic way, we embed the quarkyonic construction within relativistic mean-field (RMF) theory \cite{WALECKA1974491,BOGUTA1977413,Reinhard_1989,HOROWITZ1981503,patt1,patt2,patt3,patt4}. The RMF models provide a covariant and computationally efficient framework widely used for finite nuclei and neutron-star matter, and different parameterizations correspond to different density dependences of the mean fields and thus to different stiffness patterns at supra-saturation density \cite{estel1,patra91,Ring_1996,Dutra_2014}. Using the G3 and IOPB-I \cite{Kumar_2017,Kumar_2018} parameterizations as hadronic baselines, we generate families of quarkyonic EOSs by varying the transition density $n_t$ and the confinement scale $\Lambda_{\rm cs}$, which together determine the onset and strength of the quarkyonic crossover \cite{Dey_1,Dey_2,McLerran_2019}.

While masses, radii, and tidal deformabilities provide powerful integral constraints, they do not uniquely determine the detailed internal structure of the star \cite{Lattimer_2007}. Complementary information can be obtained from neutron-star seismology (asteroseismology), which probes the dynamical response of the star through its quasinormal modes (QNMs) \cite{Anderson_1996,obs_gw_2006}. Each mode family is sensitive to different aspects of the stellar structure and composition. Fluid-led modes (such as the $f$- and $p$-modes) are governed primarily by the bulk compressibility of matter and can couple strongly to tidal forcing and post-merger dynamics \cite{Anderson_1996,Anderson_wmode_obs_1998,merger_2015,post_merger_obs_1}.

In contrast, the spacetime-led $\omega$ modes are governed mainly by the relativistic curvature potential outside, and close to, the stellar surface \cite{Kokkotas_1992,Leins_1993,ur_wmode_2005}. They have no Newtonian counterpart, excite only weak fluid motion, and are rapidly damped, with typical damping times of order $\sim 10^{-4} \mathrm{s}$\ \cite{Anderson_1998}. Their characteristic frequencies lie in the high-frequency band, typically $\sim 5$--$20,\mathrm{kHz}$ for neutron-star models \cite{Anderson_wmode_obs_1998,Kokkotas_wmode_2004}. These modes are understood as "pure space-time modes," a concept clarified by studies using the Inverse Cowling Approximation, which neglects fluid motion \cite{Anderson_inverse_cowling}. Their existence stems from the scattering of gravitational waves off the spacetime curvature of the star, as detailed in the seminal work by Chandrasekhar and Ferrari \cite{Chandrasekhar_wmode_1}. Although this frequency range lies above the most sensitive band of current detectors, it has been argued that $\omega$ modes can be excited in dynamical scenarios such as the collapse of a neutron star to a black hole shortly before horizon formation and through the scattering of GWs by a compact star \cite{Benhar_2004,Chandrasekhar_wmode_1,gw_detectors_2019}. Moreover, future third-generation GW observatories and improved high-frequency sensitivity motivate the development of robust theoretical tools to extract the astrophysical information encoded in these spacetime modes \cite{Benhar_2004}.

A key theme in the modern asteroseismology literature is the existence of approximate universal relations linking QNM observables---typically reported through a mode frequency $f$ and a damping time $\tau$---to macroscopic stellar properties, often with only mild EOS dependence \cite{ur_wmode_2005,urfmode_2013,ur_fmode_2015,ur_fmode_2024}. Such relations are valuable because they provide a practical route from measured high-frequency ringdown features to bulk properties such as the mass, radius, compactness, and tidal deformability \cite{ur_wmode_2005}. At the same time, deviations from universality can carry information about additional microphysics (e.g., phase transitions or crossover behavior) and can therefore be used as diagnostics of the stellar interior \cite{ur_wmode_2005,obs_gw_2006,Fujimoto_gwwave_signal_2023}.

In this context, the central goal of the present work is to assess how quarkyonic microphysics imprints itself on the $\omega$ mode spectrum. We compute the complex eigenfrequencies $\omega=\omega_R+i\omega_I$, which are typically reported in terms of an oscillation frequency $f=\omega_R/2\pi$ and a damping time $\tau=1/|\omega_I|$ \cite{Anderson_1998}. To determine these quantities accurately for highly damped modes, we employ numerical techniques developed for quasinormal modes, such as those used for black hole perturbations \cite{regge_wheeler_1957,leaver_QNM_techniques_1985,Nollert_QNM_techniques_1992,Nollert_QNM_techniques_1993}. Overall, by combining a quarkyonic-RMF EOS construction with robust QNM calculations, our study aims to provide a coherent bridge between high-density microphysics, macroscopic stellar structure, and potentially observable high-frequency spacetime ringing, and to quantify how variations in $(n_t,\Lambda_{\rm cs})$ shift the fundamental and first-overtone $\omega$ mode branches in ways that may be relevant for interpreting mass-gap candidates such as GW190814 and GW230529 \cite{jeetGW230529} in a multimessenger setting \cite{Abbott_2020,Abac_2024,Fujimoto_merger_2025}.

The paper is organized as follows. Section~\ref{NM} and Section ~\ref{QM} present the nuclear model and quarkyonic-matter model within the RMF framework respectively. The cross-over transition between hadron and quark is treated in Section \ref{CRO}. Section~\ref{eqlbm} describes the equilibrium configuration of an isolated, nonrotating neutron star. In Section~\ref{Pert}, we summarize the perturbation formalism and the oscillation equations, and we briefly describe the phase--amplitude method used to compute the quasinormal mode spectrum. The results are discussed in Section~\ref{RD}, and we conclude in Section~\ref{Concl.}.
\\
\section{Construction of the Equation of State (EOS)}
\subsection{Nuclear model}
\label{NM}
This section details the models for the nucleonic, quarkyonic employing the RMF theory for the nucleonic component. The RMF formalism is a robust and versatile framework, successfully applied to describe diverse states of matter—from infinite nuclear matter and finite nuclei, including exotic nuclei near the drip lines, to the superdense matter in neutron star interiors \cite{WALECKA1974491,BOGUTA1977413,Serot1992,Ring_1996,Dutra_2014,estel1,estel2,patra91}. In this model, with a small number of parameters and the masses of the mesons, the approach reproduces the experimental data similar to the Skyrme formalism or even better throughout the mass table. Its applicability spans a vast density range, from subsaturation regimes to supra-saturation matter properties in NS cores \cite{Ring_1996,Dutra_2014}.

The model is built upon a Lagrangian density that incorporates interactions among nucleons mediated by mesons, including nonlinear self-couplings and cross-couplings \cite{WALECKA1974491,BOGUTA1977413,Serot1992,Kumar_2017, Kumar_2018,E-RMF9}. For the present study, we use the Effective RMF (E-RMF) model, which specifically includes these mesonic couplings up to the 4th order \cite{E-RMF9,E-RMF5,Kumar_2017,Kumar_2018}. The total energy density and pressure for the system, which consists of nuclear matter and leptons, are derived from this Lagrangian via the stress-energy tensor \cite{E-RMF5}. 
\label{Formalism1}
\begin{eqnarray}
\label{eq:eden}
{\cal E}_{\rm NML} & = & \sum_{i=p,n} \frac{g_s}{(2\pi)^{3}}\int_{0}^{k_{f_{i}}} d^{3}k\, \sqrt{k^{2} + M_{\rm nucl.}^{*2}}\nonumber\\
&&
+n_{b} g_\omega\,\omega+m_{\sigma}^2{\sigma}^2\Bigg(\frac{1}{2}+\frac{\kappa_{3}}{3!}\frac{g_\sigma\sigma}{M_{\rm nucl.}}+\frac{\kappa_4}{4!}\frac{g_\sigma^2\sigma^2}{M_{\rm nucl.}^2}\Bigg)
\nonumber\\
&&
 -\frac{1}{4!}\zeta_{0}\,{g_{\omega}^2}\,\omega^4
 -\frac{1}{2}m_{\omega}^2\,\omega^2\Bigg(1+\eta_{1}\frac{g_\sigma\sigma}{M_{\rm nucl.}}+\frac{\eta_{2}}{2}\frac{g_\sigma^2\sigma^2}{M_{\rm nucl.}^2}\Bigg)
 \nonumber\\
&&
 + \frac{1}{2} (n_{n} - n_{p}) \,g_\rho\,\rho
 -\frac{1}{2}\Bigg(1+\frac{\eta_{\rho}g_\sigma\sigma}{M_{\rm nucl.}}\Bigg)m_{\rho}^2
 \nonumber\\
 && 
-\Lambda_{\omega}\, g_\rho^2\, g_\omega^2\, \rho^2\, \omega^2
+\frac{1}{2}m_{\delta}^2\, \delta^{2}
\nonumber\\
 && 
+\sum_{j=e,\mu}  \frac{g_s}{(2\pi)^{3}}\int_{0}^{k_{F_{j}}} \sqrt{k^2 + m^2_{j}} \, d^{3}k,
\end{eqnarray}
and
\begin{eqnarray}
\label{eq:press}
P_{\rm NML} & = & \sum_{i=p,n} \frac{g_s}{3 (2\pi)^{3}}\int_{0}^{k_{f_{i}}} d^{3}k\, \frac{k^2}{\sqrt{k^{2} + M_{\rm nucl.}^{*2}}} \nonumber\\
&& - m_{\sigma}^2{\sigma}^2\Bigg(\frac{1}{2} + \frac{\kappa_{3}}{3!}\frac{g_\sigma\sigma}{M_{\rm nucl.}} + \frac{\kappa_4}{4!}\frac{g_\sigma^2\sigma^2}{M_{\rm nucl.}^2}\Bigg)+ \frac{1}{4!}\zeta_{0}\,{g_{\omega}^2}\,\omega^4 
\nonumber\\
&&
+\frac{1}{2}m_{\omega}^2\omega^2\Bigg(1+\eta_{1}\frac{g_\sigma\sigma}{M_{\rm nucl.}}+\frac{\eta_{2}}{2}\frac{g_\sigma^2\sigma^2}{M_{\rm nucl.}^2}\Bigg)
\nonumber\\
&&
+ \frac{1}{2}\Bigg(1+\frac{\eta_{\rho}g_\sigma\sigma}{M_{\rm nucl.}}\Bigg)m_{\rho}^2\,\rho^{2}-\frac{1}{2}m_{\delta}^2\, \delta^{2}+\Lambda_{\omega} g_\rho^2 g_\omega^2 \rho^2 \omega^2
\nonumber\\
&&
+\sum_{j=e,\mu}  \frac{g_s}{3(2\pi)^{3}}\int_{0}^{k_{F_{j}}} \frac{k^2}{\sqrt{k^2 + m^2_{j}}} \, d^{3}k.
\end{eqnarray}
\noindent
Where $g_s$ and $M_{nucl.}$ represent the spin degeneracy and mass of the nucleon. The $m_\sigma$, $m_\omega$, $m_\rho$, and $m_\delta$ are the masses, and $g_\sigma$, $g_\omega$, $g_\rho$, and $g_\delta$ are the coupling constants for the $\sigma$, $\omega$, $\rho$, and $\delta$ mesons respectively. Other couplings, such as $\kappa_3$, $\kappa_4$, $\zeta_0$ are for the self-interactions, and $\eta_1$, $\eta_2$, $\eta_\rho$, and $\Lambda_\omega$ are the cross-couplings between mesons \cite{ERMF6, Serot1992, ERMF7, E-RMF8, E-RMF9, E-RMF5, Kumar_2018}.\\
\subsection{Quarkyonic model}
\label{QM}
We briefly review the quarkyonic model introduced by McLerran and Reddy \cite{McLerran_2019}. In this framework, it is proposed that in the inner core of neutron stars—where densities exceed several times the nuclear saturation density—nucleons begin to dissolve into quark degrees of freedom. A defining aspect of this model is its impact on the mass–radius relation, predicting higher maximum masses due to the stiffened EOS relative to purely nucleonic matter. The transition to the quarkyonic phase is characterized by a rapid increase in pressure, arising from the population of low-momentum quark states once the baryon density crosses a critical threshold  transition density $n_t$. Consequently, low-momentum states are treated as quark degrees of freedom, whereas high-momentum states near the Fermi surface remain nucleonic. Since momenta near the Fermi surface are of the order of the QCD confinement scale $\Lambda_{\rm cs}$, they favor the formation of bound quark states, i.e., nucleons.

The original schematic model of McLerran and Reddy is subsequently extended by Zhao and Lattimer \cite{tinaki_ur}. Their formulation incorporates beta equilibrium and charge neutrality conditions in quarkyonic matter. In their approach, nucleon interactions are described through density-dependent potentials calibrated to properties of uniform nuclear matter. In contrast, our model employs relativistic mesonic mean fields instead of density-dependent nucleonic potentials. Furthermore, chemical equilibrium between nucleons and quarks is imposed to relate the Fermi momenta of nucleons ($k_{f_{n,p}}$) and quarks ($k_{f_{u,d}}$), which is a key feature of the modified quarkyonic description. Due to the shell structure of nucleons in momentum space, nucleons occupy a finite Fermi shell bounded between a minimum momentum $k_{0_{(n,p)}}$ and a maximum momentum $k_{f_{(n,p)}}$. The up and down quarks occupy Fermi seas characterized by momenta $k_{f_u}$ and $k_{f_d}$, respectively.

The conservation of baryon number density is expressed as \cite{tinaki_ur}
\begin{eqnarray}
n &=& n_n + n_p + \frac{n_u+n_d}{3} \nonumber\\
&=& \int_{k_{0_n}}^{k_{f_n}} \frac{g_s}{2\pi^2} k^2 dk \;+\; \int_{k_{0_p}}^{k_{f_p}} \frac{g_s}{2\pi^2} k^2 dk \;\\&&+\; \frac{1}{3}\left( \int_0^{k_{f_u}} \frac{g_s}{2\pi^2} k^2 dk + \int_0^{k_{f_d}} \frac{g_s}{2\pi^2} k^2 dk \right) \nonumber\\
&=& \frac{g_s}{6\pi^2}\bigg[(k_{f_{n}}^3-k_{0_{n}}^3)+(k_{f_{p}}^3-k_{0_{p}}^3)+\frac{(k_{f_{u}}^3+k_{f_{d}}^3)}{3}\bigg].
\end{eqnarray}
The condition of electric charge neutrality among nucleons, quarks, and leptons is given by
\begin{eqnarray}
n_p + \frac{2n_{u}}{3} -  \frac{n_{d}}{3} = n_{e^{-}} + n_{\mu}.
\end{eqnarray}
The lower bound of the nucleon Fermi momentum is defined as
\begin{eqnarray}
 k_{0(n,p)} = (k_{f_{(n,p)}}-k_{t_{(n,p)}})\bigg[1+ \frac{\Lambda_{cs}^2}{k_{f_{(n,p)}}k_{t_{(n,p)}}}\bigg].
\end{eqnarray}
where $\Lambda_{cs}$ represents the confinement scale, and $k_{t_{(n,p)}} = (3\pi^2 n_{t_{(n,p)}})^{1/3}$ corresponds to the Fermi momentum at the transition density $n_t = n_{t_n} + n_{t_p}$.
The requirement of strong interaction equilibrium ensures that, at a given baryon density, the system minimizes its total energy. This condition translates into chemical equilibrium between nucleons and quarks, expressed as \cite{tinaki_ur}
\begin{eqnarray}
\mu_n &=& \mu_u + 2\mu_d, \\
\mu_p &=& 2\mu_u + \mu_d,
\end{eqnarray}
where $\mu_n$, $\mu_p$, $\mu_u$, and $\mu_d$ denote the chemical potentials of neutrons, protons, up quarks, and down quarks, respectively.

In addition, beta equilibrium under charge neutrality further constrains the system and establishes equilibrium among nucleons and leptons \cite{Glendenning,tinaki_ur}:
\begin{eqnarray}
\mu_{n} &=& \mu_{p} + \mu_{e^{-}}, \nonumber \\
\mu_{\mu} &=& \mu_{e^{-}}.
\end{eqnarray}
An important feature of the model is that the effective masses of up and down quarks are not treated as independent parameters. Instead, they are determined from beta equilibrium conditions at the transition density $n_t$, leading to
\begin{eqnarray}
 m_{u} = \frac{2}{3} \mu_{t_{p}} -  \frac{1}{3} \mu_{t_{n}}, \quad
 m_{d} = \frac{2}{3} \mu_{t_{n}} -  \frac{1}{3} \mu_{t_{p}},
\end{eqnarray}
where $\mu_{t_n}$ and $\mu_{t_p}$ are the neutron and proton chemical potentials evaluated at the transition density in beta-equilibrated matter consisting of nucleons and leptons.
Assuming quarks behave as a non-interacting Fermi gas, their energy density ${\cal E}_{\rm QM}$ and pressure $P_{\rm QM}$ are given by \cite{tinaki_ur}
\begin{eqnarray}
{\cal E}_{\rm QM} = \sum_{j=u,d}\frac{g_s N_c}{(2\pi)^3}\int_0^{k_{f_{j}}} k^2 \sqrt{k^2 + m_{j}^2}\, d^3k,
\end{eqnarray}
\begin{eqnarray}
P_{\rm QM} = \mu_{u} n_{u} + \mu_{d} n_{d} - {\cal E}_{\rm QM}.
\end{eqnarray}

The present study does not attempt a strict first-principles realization of quarkyonic matter in the original sense proposed by McLerran and Reddy. Instead, we construct a phenomenological, quarkyonic-inspired crossover EOS within a RMF framework and investigate its ability to sustain massive compact stars. A central feature of quarkyonic matter is the rapid enhancement of the speed of sound near the transition density. This characteristic leads to a significant stiffening of the EOS, thereby allowing the support of heavier neutron stars. Notably, this essential behavior is also reproduced in our model.

\subsection{Quarkyonic-inspired crossover equation of state}
\label{CRO}
In the present study, quarkyonic-inspired equations of state are formulated using a smooth crossover construction rather than a first-order phase transition \cite{jeetGW230529}. The methodology follows the interpolated equation of state scheme originally proposed by Masuda et al.~\cite{Masuda2013} and later adopted in the quarkyonic context by Han et al.~\cite{sophia_han_19}. Unlike conventional Maxwell or Gibbs constructions, this framework allows for a continuous evolution from hadronic to quarkyonic degrees of freedom across a finite density interval.
Within this description, the parameter $n_t$ characterizes the density scale at which quarkyonic features begin to emerge, while the smoothness of the transition is governed by a width parameter $\Gamma$. As a result, the equation of state is not constrained to match the purely nucleonic description up to a single, well-defined transition density, but instead undergoes a gradual crossover.
The crossover is implemented by interpolating the pressure as a function of
baryon number density,
\begin{equation}
P(n) = P_{\rm NML}(n)\,f_{-}(n) + P_{\rm QM}(n)\,f_{+}(n),
\label{eq:Pinterp}
\end{equation}
which is equivalent to the prescription given in Eq.~(19) of
Ref.~\cite{sophia_han_19}. Here, $P_{\rm NML}(n)$ denotes the pressure obtained from the purely hadronic RMF equation of state, while $P_{\rm QM}(n)$ corresponds to the quarkyonic-inspired sector. The weighting functions are defined such that
\begin{equation}
f_{-}(n) + f_{+}(n) = 1,
\qquad
f_{\pm}(n) =
\frac{1}{2}
\left[
1 \pm \tanh\!\left(\frac{n-n_t}{\Gamma}\right)
\right],
\label{eq:weights}
\end{equation}
ensuring a smooth interpolation across the crossover region. The parameter
$\Gamma$ controls the density range over which the transition occurs and is chosen phenomenologically to maintain numerical stability and smooth thermodynamic behavior.
A direct interpolation of the pressure necessitates a careful reconstruction of the energy density in order to preserve thermodynamic consistency. Following the procedure outlined by Masuda et al.~\cite{Masuda2013} and explicitly implemented in Ref.~\cite{sophia_han_19}, the total energy density in the crossover region is written as
\begin{equation}
{\cal E}(n)
=
{\cal E}_{\rm NML}(n)\,f_{-}(n)
+
{\cal E}_{\rm QM}(n)\,f_{+}(n)
+
\Delta{\cal E}(n),
\label{eq:energy}
\end{equation}
which corresponds to Eq.~(21) of Ref.~\cite{sophia_han_19}. The correction term
$\Delta{\cal E}(n)$ arises due to the explicit density dependence of the
interpolation functions and takes the form
\begin{equation}
\Delta{\cal E}(n)
=
n \int_{n_t}^{n}
dn'\,
\frac{{\cal E}_{\rm NML}(n') - {\cal E}_{\rm QM}(n')}
{n'}\, g(n'),
\label{eq:deltaeps}
\end{equation}
where
\begin{equation}
g(n') = \frac{d f_{+}(n')}{dn'}
      = \frac{1}{2\Gamma}
        \mathrm{sech}^2\!\left(\frac{n'-n_t}{\Gamma}\right).
\end{equation}
This term is not introduced phenomenologically but follows directly from enforcing the thermodynamic relation
\begin{equation}
P = n^2 \frac{\partial}{\partial n}\left(\frac{\varepsilon}{n}\right),
\end{equation}
thereby ensuring internal consistency of the interpolated equation of state. In our earlier work, we employed this method, which has proven to be effective in describing multimessenger observational constraints \cite{Dey_1,Dey_2}. 

\section{The Equilibrium State }
\label{eqlbm}

The equilibrium configuration of a non-rotating neutron star is determined by solving the Einstein field equations for a self-gravitating, spherically symmetric perfect fluid. The solution of which is the TOV equations given as \cite{PhysRev.55.364,PhysRev.55.374,Shapiro_1983}:
\begin{align}
\frac{dm}{dr} &= 4\pi r^{2} \mathcal{E}(r), \label{eq:mass_eq} \\
\frac{dP}{dr} &= -\frac{G}{r^{2}} \left[\mathcal{E}(r) + P(r)\right] \left[m(r) + 4\pi r^{3} P(r)\right] e^{\lambda(r)}, \label{eq:tov_eq}
\end{align}
\noindent
where the mass function $m(r)$ represents the mass enclosed within radius $r$ with $e^{-\lambda(r)}=1-\frac{2m(r)}{r}$.

Equation~\eqref{eq:mass_eq} defines the gravitational mass, while Equation~\eqref{eq:tov_eq} governs hydrostatic equilibrium. The term $\left[m(r) + 4\pi r^{3} P(r)\right]$ reveals pressure's role as a gravitational source in general relativity, a fundamental departure from Newtonian theory. The factor $e^{\lambda(r)}$ accounts for spacetime curvature effects, which become significant in the high-density core. The system is closed by an equation of state $P(\mathcal{E})$. Integration proceeds from the center ($r=0$) with boundary conditions $m(0) = 0$ and a central pressure $P(0) = P_c$, outward to the stellar surface $R$ defined by $P(R) = 0$. The total gravitational mass is then $M = m(R)$.

\section{Perturbed state }
The neutron star undergoes coupled perturbations in its interior matter distribution and exterior spacetime, where fluctuations of the relativistic fluid are dynamically linked to spacetime curvature through Einstein’s field equations. These perturbations encode information about the dense-matter equation of state and the star’s internal composition, while the exterior response governs the propagation of gravitational radiation. We briefly outlined the interior and exterior regions of the NS below.
\label{Pert}
 \subsection{Interior region of the Neutron star}

The interior of a nonrotating neutron star is characterized by two fundamental degrees of freedom: the matter perturbations, describing oscillations of the dense nuclear fluid, and the spacetime perturbations, which represent the associated ripples in the gravitational field \cite{Throne_1967,Lindblom_1983}. This coupling is essential for modeling stellar oscillations that can emit gravitational waves. The even-parity perturbations in the Regge--Wheeler gauge are described by the metric~\cite{Lindblom_1983}.
\begin{equation}
\begin{aligned}
ds^{2} = & -e^{\nu}\left(1 + r^{\ell} H_{0} Y_{\ell m} e^{i\omega t}\right) dt^{2} \\
          & + e^{\lambda}\left(1 - r^{\ell} H_{0} Y_{\ell m} e^{i\omega t}\right) dr^{2} \\
          & + r^{2}\left(1 - r^{\ell} K Y_{\ell m} e^{i\omega t}\right) d\Omega^{2} \\
          & - 2i \omega r^{\ell+1} H_{1} Y_{\ell m} e^{i\omega t} \, dt \, dr,
\end{aligned}
\end{equation}
where \(H_0\), \(H_1\), and \(K\) are radial functions representing the metric perturbations, and \(\omega\) is the complex oscillation frequency whose real part gives the mode frequency and imaginary part the damping rate due to gravitational wave emission. The $\nu$ can be calculated using the relation $\frac{d\nu}{dr}=2~{}\frac{(M+4\pi r^{3}p)}{r(r-2M)}$. The fluid perturbations are described by the Lagrangian displacement vector, which characterizes how fluid elements move during oscillations \cite{Throne_1967,Lindblom_1983}:
\begin{equation}
\begin{aligned}
\xi^r      &= r^{\ell-1} e^{-\lambda/2} W Y_{\ell m} e^{i\omega t}, \\
\xi^\theta &= -r^{\ell-2} V \partial_\theta Y_{\ell m} e^{i\omega t}, \\
\xi^\phi   &= -\frac{r^{\ell-2}}{\sin^2\theta} V \partial_\phi Y_{\ell m} e^{i\omega t},
\end{aligned}
\end{equation}
\noindent
where \(W\) and \(V\) represent the radial and the horizontal (tangential) displacement, respectively. The $Y_{\ell m}(\theta, \phi)=Y_{\ell m}(e^{i\omega t})$ represents the spherical harmonics. 

To formulate a well-posed eigenvalue problem for \(\omega\), one must avoid singularities that arise in the perturbation equations. This is achieved by introducing a new variable \(X\), which is related to the Lagrangian pressure variation:
\begin{equation}
\Delta p = -r^\ell e^{-\nu/2} X Y_{\ell m} e^{i\omega t}.
\end{equation}
The system is then reduced to four first-order differential equations for the variables \(H_1\), \(K\), \(W\), and \(X\) \cite{Lindblom_1983,detweiler85}, which together describe the coupled fluid-spacetime system. The evolution of these variables is governed by\cite{detweiler85}:
\begin{equation}
\begin{aligned}
r \frac{dH_1}{dr} &= -\left[\ell + 1 + 2b e^\lambda + 4\pi r^2 e^\lambda (p - \epsilon)\right] H_1 \\
                  & \quad + e^\lambda \left[H_0 + K - 16\pi (\epsilon + p) V\right], \\ 
r \frac{dK}{dr} &= H_0 + (\ell + 1) H_1 + \left[e^\lambda Q - \ell - 1\right] K \\
                & \quad - 8\pi (\epsilon + p) e^{\nu/2} W, \\
r \frac{dW}{dr} &= -(\ell + 1)\left[W + \ell e^{\nu/2} V\right] \\
                & \quad + r^2 e^{\lambda/2} \left[\frac{X}{(\epsilon + p)C_s^2} e^{-\nu/2} + \frac{H_0}{2} + K\right], \\
r \frac{dX}{dr} &= -\ell X + \frac{\epsilon + p}{2} e^{\nu/2} \Bigg\{ (3e^\lambda Q - 1) K \\
                & \quad - \frac{4(\ell^2 + \ell - 2) e^\lambda Q}{r^2} V + (1 - e^\lambda Q) H_0 \\
                & \quad + \left(r^2 \omega^2 e^{-\nu} + \ell^2 + \ell - 2\right) H_1 \\
                & \quad + \bigg[2\omega^2 e^{\lambda/2 - \nu} - 8\pi (\epsilon + p) e^{\nu/2} \\
                & \quad \quad + r^2 \frac{d}{dr} \left( \frac{e^{-\nu/2} g}{r^2} \right) \bigg] W \Bigg\},
\end{aligned}
\end{equation}
where \(Q = M + 4\pi r^3 p\) represents the gravitational mass including pressure contributions, \(b = GM/r\) is the compactness, and \(C_s\) is the adiabatic sound speed. The first equation governs the evolution of the metric perturbation \(H_1\), which is coupled to fluid variables through \(V\) and \(W\). The second equation describes the curvature perturbation \(K\), which sources and is sourced by both metric and fluid terms. The third equation determines the radial fluid displacement \(W\), driven by pressure perturbations \(X\) and metric variations. The fourth equation controls the pressure perturbation variable \(X\), which is influenced by all other variables and encodes the fluid's response to spacetime curvature changes.

The remaining metric function \(H_0\) and fluid function \(V\) are not independent but are determined algebraically by constraint equations that ensure consistency with Einstein's field equations\cite{tinaki_ur}:
\begin{equation}
\begin{aligned}
H_0 &= \frac{1}{2b + \ell + Q} \Big\{ 8\pi r^2 e^{-\nu/2} X \\
    &\quad - \left[(\ell^2 + \ell - 1) Q - \omega^2 r^2 e^{-(\nu + \lambda)}\right] H_1 \\
    &\quad + \left[\ell(\ell + 1) - \omega^2 r^2 e^{-\nu} - Q(Q e^\lambda - 1)\right] K \Big\}, \\
V &= \frac{e^{\nu/2}}{\omega^2} \left[ \frac{X}{\epsilon + p} - \frac{Q}{r^2} e^{(\nu + \lambda)/2} W - e^{\nu/2} \frac{H_0}{2} \right].
\end{aligned}
\end{equation}
These constraints reduce the number of dynamical degrees of freedom from four to two, consistent with the wave-like nature of the perturbations where only two polarizations of gravitational waves are possible.

The boundary conditions ensure physical regularity at the center (\(r = 0\)) and a free surface at the stellar boundary (\(r = R\)):
\begin{equation}
\begin{aligned}
W(0) &= 1, \\
K(0) &= H_0(0), \\
H_1(0) &= \frac{\ell K(0) + 8\pi (\epsilon_0 + p_0) W(0)}{\ell(\ell + 1)}, \\
X(0) &= (\epsilon_0 + p_0) e^{\nu_0/2} \left[ \frac{4\pi}{3} (\epsilon_0 + 3p_0) - \frac{\omega^2}{\ell} e^{-\nu_0} \right] W(0) \\
      &\quad + \frac{K(0)}{2},
\end{aligned}
\end{equation}
with the surface condition \(X(R) = 0\) ensuring vanishing Lagrangian pressure perturbation. This formulation captures the complete physics of nonradial oscillations in relativistic stars, providing the foundation for calculating quasinormal modes and their gravitational wave signatures.

\subsection{The exterior region of the NS}

In the vacuum region outside a neutron star, where matter perturbations vanish, the dynamics are governed solely by the degrees of freedom of the spacetime itself. The two independent metric perturbations, denoted by \( H_1 \) and \( K \), can be combined into a single master variable satisfying a second-order wave equation known as the Zerilli equation \cite{fackerell71,zerilli_1970}. This equation describes the propagation of gravitational waves in the exterior spacetime of the neutron star, defined as,

\begin{equation}
\left[ \frac{d^2}{dr_*^2} + \omega^2 - V_{\rm Z}(r) \right] Z(r) = 0,
\label{zerilli_eq}
\end{equation}
\noindent
where \( Z(r) \) is the Zerilli function, \( \omega \) is the complex frequency of the perturbation, and \( V_{\rm Z}(r) \) is the effective potentialgiven by,

\begin{equation}
\begin{aligned}
V_{\rm Z}(r) &= \frac{2\left(1 - \frac{2M}{r}\right)}{r^3(nr + 3M)^2} \\
&\quad \times \left[ n^2(n+1)r^3 + 3n^2 M r^2 + 9n M^2 r + 9M^3 \right],
\end{aligned}
\end{equation}
\noindent
with \( n = \frac{1}{2}(l-1)(l+2) \), where \( l \) is the angular quantum number, and \( M \) represents the total mass of the neutron star. The Zerilli potential \( V_{\rm Z}(r) \) is of fundamental importance as it encapsulates the entire influence of the curved spacetime on the propagating gravitational waves. Its shape creates a potential barrier surrounding the neutron star, which is peaked just outside the stellar radius. This barrier is responsible for several key physical phenomena: it partially traps gravitational radiation, leading to the existence of long-lived quasinormal modes; it filters and scatters incoming waves, determining the reflection and transmission coefficients for gravitational radiation; and its height and width directly influence the damping times and frequencies of the oscillations. The potential vanishes both at the stellar surface (approximated by the Schwarzschild radius in this exterior solution) and at spatial infinity, ensuring that wave solutions become simple plane waves in these asymptotic regions \cite{Benhar_2004}.

The tortoise coordinate \( r_* \) plays a crucial role in the analysis of wave propagation around neutron stars. It is defined by the transformation:
\begin{equation}
\frac{d r_*}{d r} = \frac{1}{1 - \frac{2M}{r}},
\end{equation}
This coordinate transformation is particularly significant for neutron star physics as it regularizes the wave equation at the surface and beyond. While for black holes the tortoise coordinate maps the event horizon to negative infinity, for neutron stars (which lack an event horizon) it serves to simplify the wave equation and facilitate the implementation of boundary conditions \cite{Benhar_2004}. The coordinate stretches the space around the compact object, providing a conformally flat background that is essential for cleanly separating incoming and outgoing wave solutions at infinity and ensuring proper treatment of the wave propagation in the strong-field region near the neutron star surface.

The solutions to the Zerilli equation represent gravitational wave modes in the neutron star's exterior. For general frequencies \( \omega \), the physical solution consists of a mixture of outgoing and ingoing waves at spatial infinity. The quasinormal modes correspond to specific discrete, complex frequencies \( \omega_n \) that satisfy purely outgoing wave conditions at infinity. These modes characterize the natural oscillation frequencies of the neutron star spacetime, representing damped vibrations where gravitational wave emission carries energy away from the system. The real part of \( \omega_n \) gives the oscillation frequency, while the imaginary part determines the damping rate due to gravitational wave emission \cite{Kokkotas_1992,Benhar_2004}. 
Here we use the phase amplitude method to find out the quasinormal modes by solving the equation (\ref{eq:A6})
whose details are shown in the Appendix \ref{appendix}.

\section{Results and discussions}
\label{RD}
In this section we present the equation of state, speed of sound and mass-radius relation for quarkyonic stars. We proceed towards  the spectrum analysis of $\omega$ modes of quarkyonic stars and discuss the computed complex eigenfrequencies for the quarkyonic stars. Then we focus on the fundamental ($\omega_1$) and first-overtone ($\omega_2$) modes, analyzing how their oscillation frequencies and damping times vary with stellar structure and model parameters. Finally, we examine universal relations connecting the complex mode frequencies to bulk properties such as compactness and we assess the results of how well the quarkyonic stars  follow these trends.
  
\subsection{Equation of state, Speed of sound, and Mass-Radius}
\begin{figure}[h!]
        \centering
        \includegraphics[width=1\linewidth]{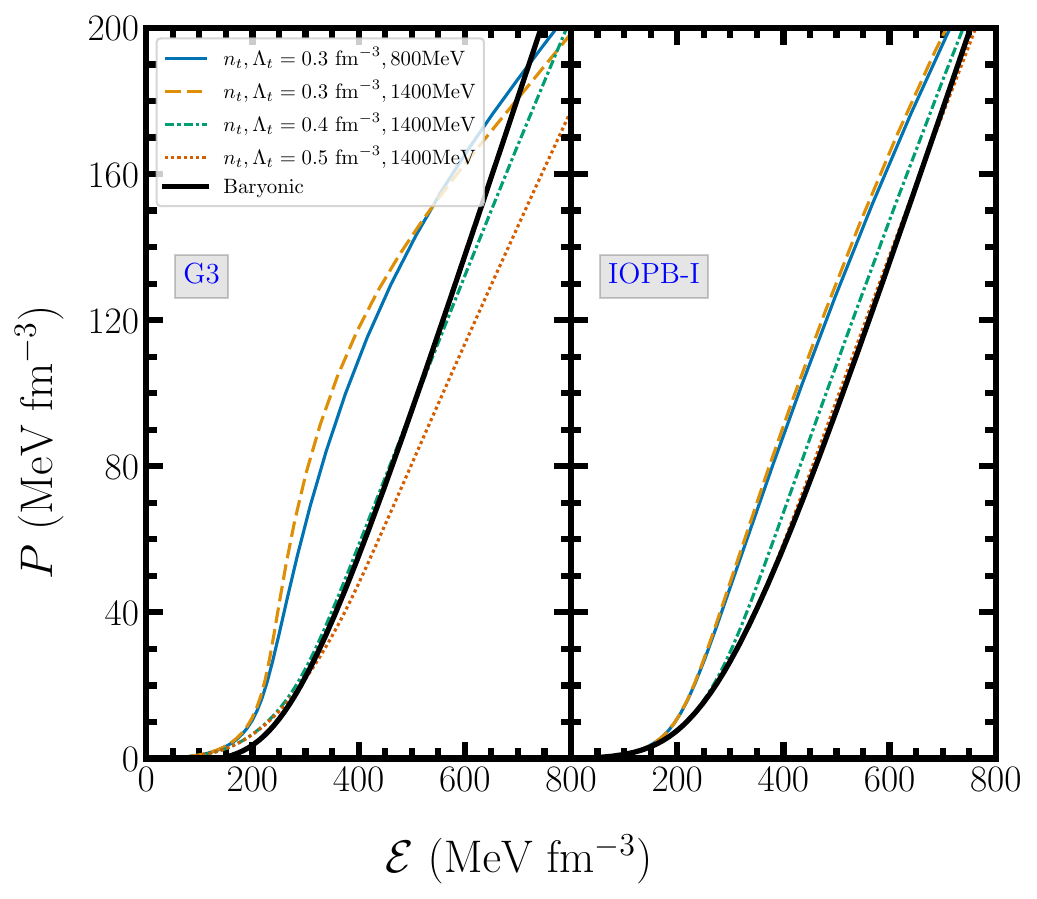}
        \caption{Equation of state showing pressure $P$ as a function of energy density $\mathcal{E}$ for quarkyonic matter constructed within the relativistic mean-field framework using G3 (left panel) and IOPB-I (right panel) parameter sets. Different curves correspond to varying transition densities $n_t$ and confinement scales $\Lambda_{cs}$. The black solid line represents the purely baryonic EOS.}
        \label{fig:eos}
    \end{figure}

\begin{figure}[h!]
        \centering
        \includegraphics[width=1\linewidth]{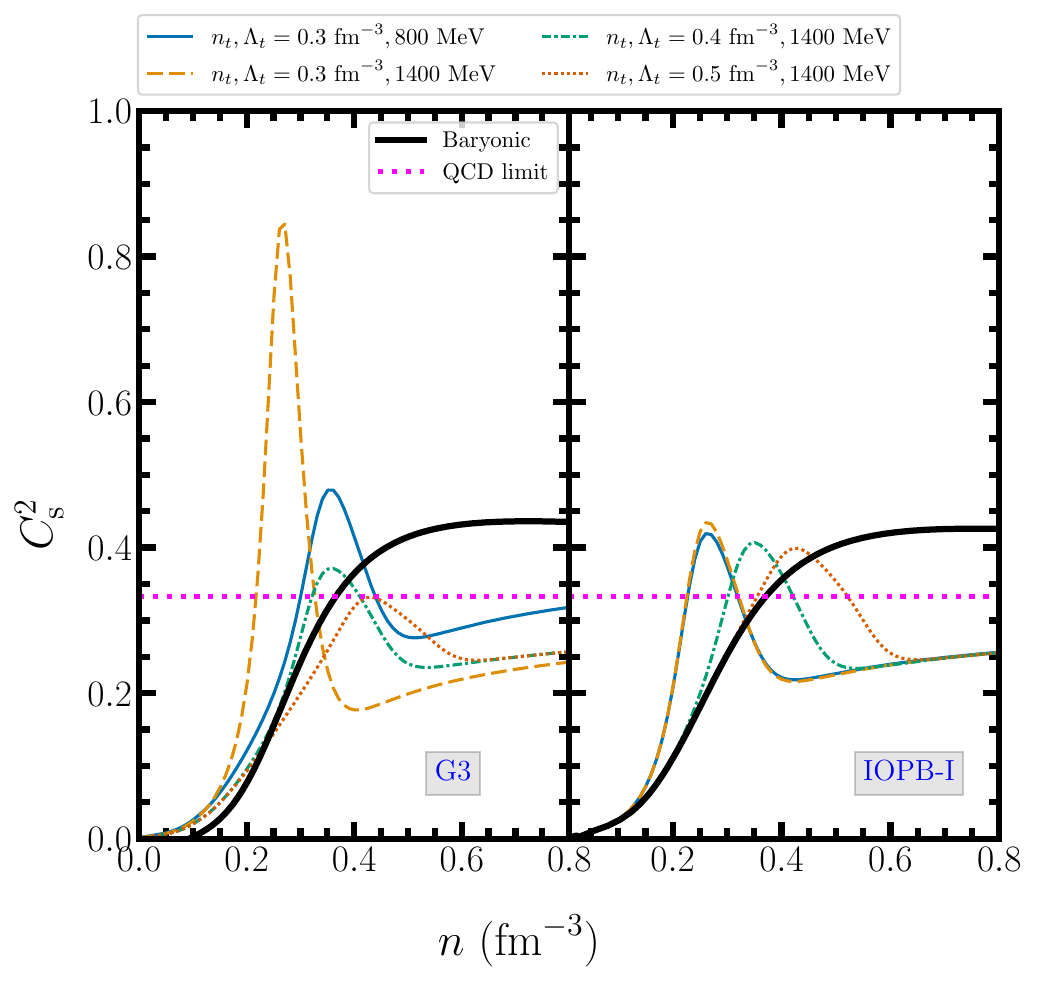}
        \caption{Squared sound speed $C_s^2$ as a function of baryon density $n$ for baryonic and quarkyonic stars using G3 (left panel) and IOPB-I (right panel) interactions. Different line styles correspond to variations in transition density $n_t$ and confinement scale $\Lambda_{cs}$. The horizontal dashed line at $C_s^2 = 1/3$ represents the conformal (QCD) limit.}
        \label{fig:sos}
    \end{figure}

\begin{figure}[h!]
        \centering
        \includegraphics[width=1\linewidth]{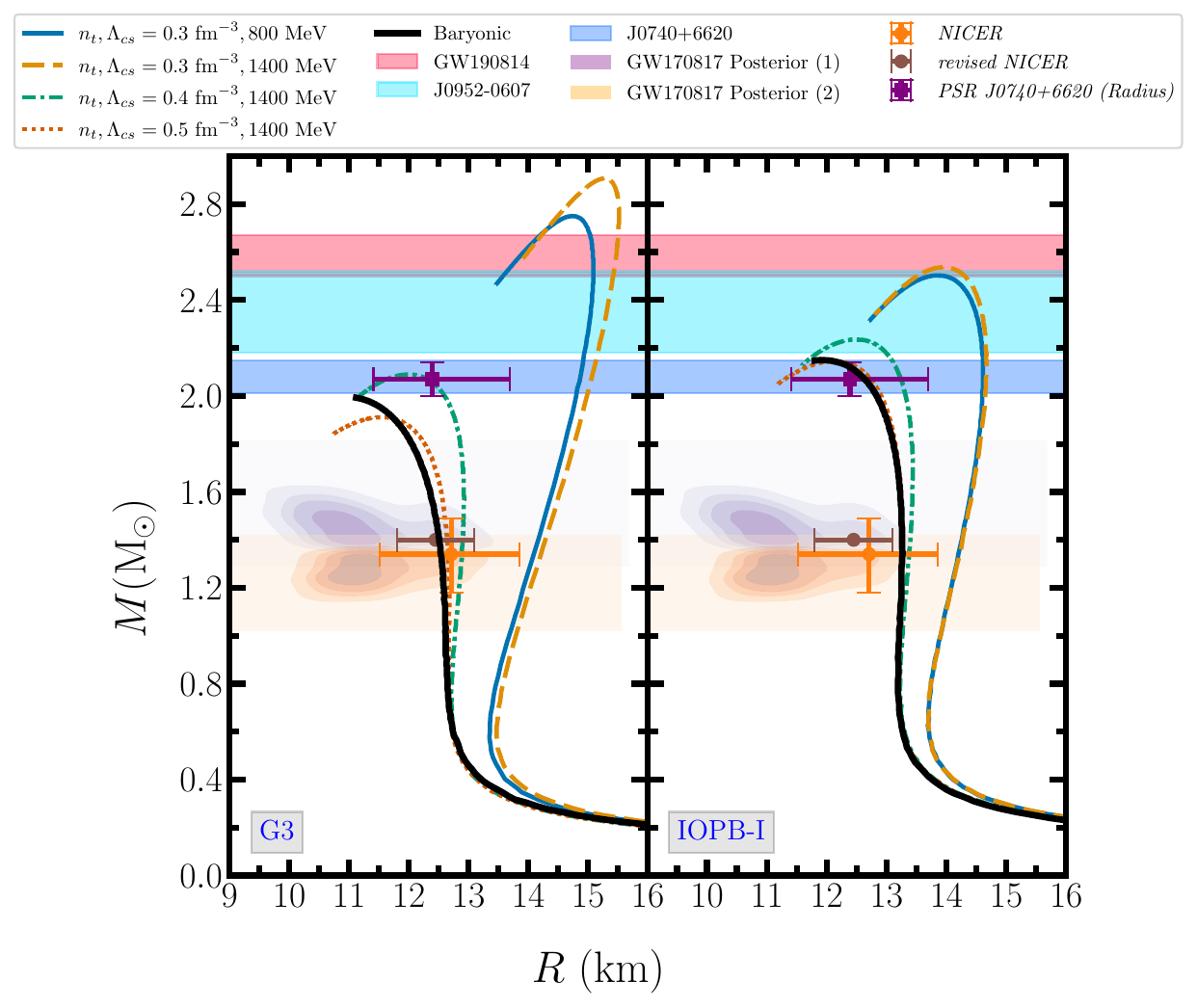}
        \caption{Mass-radius relations for baryonic and quarkyonic neutron stars obtained using G3 (left panel) and IOPB-I (right panel) parameter sets for different transition densities $n_t$ and confinement scales $\Lambda_{cs}$. The black curve denotes the purely baryonic case. Shaded regions represent observational constraints: GW190814 (red), PSR J0952--0607 (cyan), PSR J0740+6620 (blue), and GW170817 (yellow band). Error bars correspond to NICER measurements and their revised analyses.}
        \label{fig:mr}
    \end{figure}

Figures \ref{fig:eos}, \ref{fig:sos}, \ref{fig:mr} collectively illustrate the impact of quarkyonic degrees of freedom on the EOS, the corresponding speed of sound ($C_s^2=\frac{\partial P}{\partial \cal{E}}$), and the global mass--radius structure of compact stars. As shown in Fig. \ref{fig:eos}, the inclusion of quarkyonic matter leads to a pronounced stiffening of the EOS compared to the purely baryonic case. In particular, configurations with lower transition density (e.g., $n_t = 0.3~\mathrm{fm}^{-3}$) exhibit a rapid rise in pressure with energy density, reflecting the early onset of quark degrees of freedom. This stiffening behavior is a direct consequence of the population of quark states in the inner core, which enhances the pressure support at high densities. The effect is more prominent in configurations such as $(n_t, \Lambda_{cs}) = (0.3~\mathrm{fm}^{-3}, 800~\mathrm{MeV})$, where the quarkyonic contribution dominates over a larger density range, leading to a significantly stiffer EOS.

This behavior is further clarified in Fig. \ref{fig:sos} through the density dependence of the squared speed of sound $C_s^2$. The onset of quarkyonic matter is marked by a rapid increase in $C_s^2$ in the core region, often exceeding the conformal limit $C_s^2 = 1/3$, and reaching values as high as $C_s^2 \sim 0.5$--$0.8$ for lower transition densities. This sharp increase indicates a strong stiffening of the EOS in the quarkyonic regime. However, as the transition density $n_t$ increases, the onset of quarkyonic behavior is delayed to higher densities, resulting in a reduced quark fraction in the stellar interior. Consequently, the EOS becomes relatively softer in the density range relevant for neutron star cores, and the enhancement in $C_s^2$ is correspondingly suppressed. The role of the confinement scale $\Lambda_{cs}$ is comparatively subtle: for a fixed $n_t$, increasing $\Lambda_{cs}$ leads to a slightly higher quark content in the core, which marginally enhances the stiffness and modifies the sound speed profile, although the overall effect remains moderate.

The combined effect of these features is reflected in the mass--radius relations shown in Fig. \ref{fig:mr} The stiffening of the EOS at lower transition densities leads to an increase in the maximum supported mass, with configurations such as $(n_t = 0.3~\mathrm{fm}^{-3})$ reaching $M_{\mathrm{max}} \gtrsim 2.7$--$2.9\,M_\odot$, depending on the RMF parameter set. In contrast, increasing the transition density reduces the quark contribution in the core, making the stellar matter more baryon-dominated and thereby decreasing the maximum mass (e.g., down to $\sim 2.0\,M_\odot$ or below for $n_t = 0.5~\mathrm{fm}^{-3}$). The influence of the confinement scale is again subdominant but non-negligible: for a fixed transition density, larger values of $\Lambda_{cs}$ slightly increase the quark fraction, resulting in a marginal enhancement of the maximum mass and radius. Overall, these results demonstrate that the interplay between transition density and confinement scale governs the degree of EOS stiffening, which in turn controls the speed of sound profile and ultimately determines the global stellar properties.



\subsection{Spectrum analysis of $\omega$ mode}
\label{Spectrum}
\begin{figure}[h!]
        \centering
        \includegraphics[width=1\linewidth]{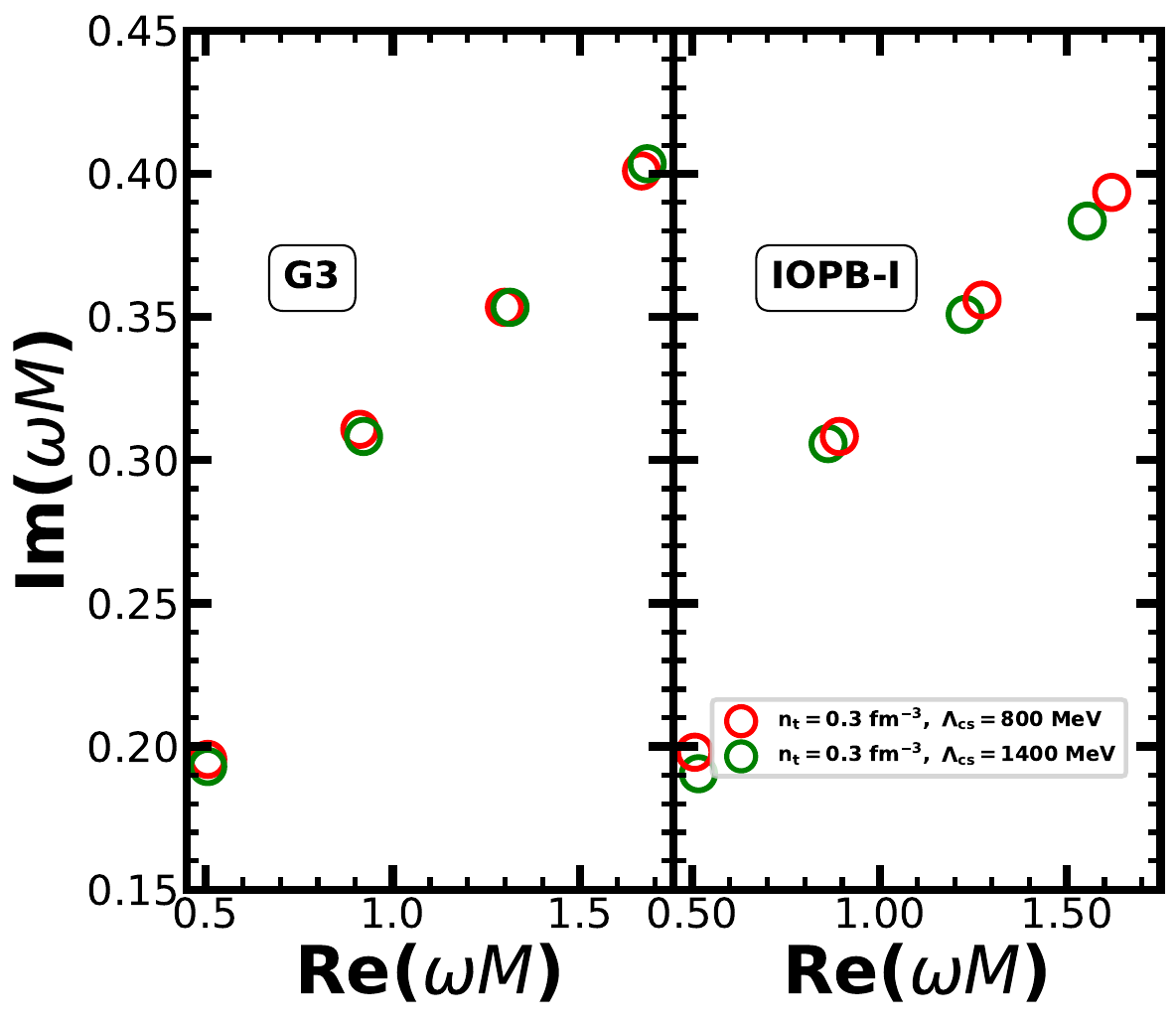}
        \caption{The dimensionless $\omega$ mode spectra ($\omega_1$, $\omega_2$, $\omega_3$, $\omega_4$) for G3 (left panel) and IOPB-I (right panel) parameter sets for quarkyonic star.}
        \label{fig:spectrum_confinement}
    \end{figure}

In this subsection we discuss the $\omega$ mode spectrum for quarkyonic star.
Figure \ref{fig:spectrum_confinement} displays the dimensionless  $\omega$ mode quasinormal spectrum for canonical stellar configurations constructed with the G3 (left panel) and IOPB-I (right panel) RMF parameter sets, highlighting the sensitivity of spacetime-led oscillations to the quarkyonic confinement scale $\Lambda_{\rm cs}$. Each point represents a complex eigenfrequency $\omega=\omega_R+i\omega_I$ (shown in a scaled, dimensionless form), where the real part sets the oscillation frequency and the imaginary part quantifies the radiative damping due to gravitational-wave emission, with a characteristic damping time scale $\tau\sim 1/|\omega_I|$. The plotted sequences form an approximately regular ladder of overtones, a characteristic feature of curvature (spacetime) modes governed by wave propagation in an effective relativistic potential barrier outside and near the stellar surface. The near-uniform spacing reflects that successive overtones correspond to progressively larger phase accumulation between the stellar surface and the peak of the exterior barrier. Varying $\Lambda_{\rm cs}$ which controls the momentum-space thickness of the nucleonic shell and how the EOS stiffens across the quarkyonic crossover. This systematically shifts the entire spectrum, most prominently in $\omega_R$. In particular, increasing $\Lambda_{\rm cs}$ pushes the spectrum toward lower $\omega_R$. This indicates that the characteristic curvature ringing becomes slower when the quarkyonic sector is modified in this direction. This is consistent with the expectation that $\omega$-modes respond primarily to changes in the global compactness profile rather than to fluid compressibility alone. The comparison between the two panels underscores that the magnitude of these shifts is EOS dependent. The G3 and IOPB-I realize different radial distributions of density and pressure for the same mass scale, which alters the curvature potential and hence the complex QNM spectrum.\\

\begin{figure}[h!]
        \centering
        \includegraphics[width=1\linewidth]{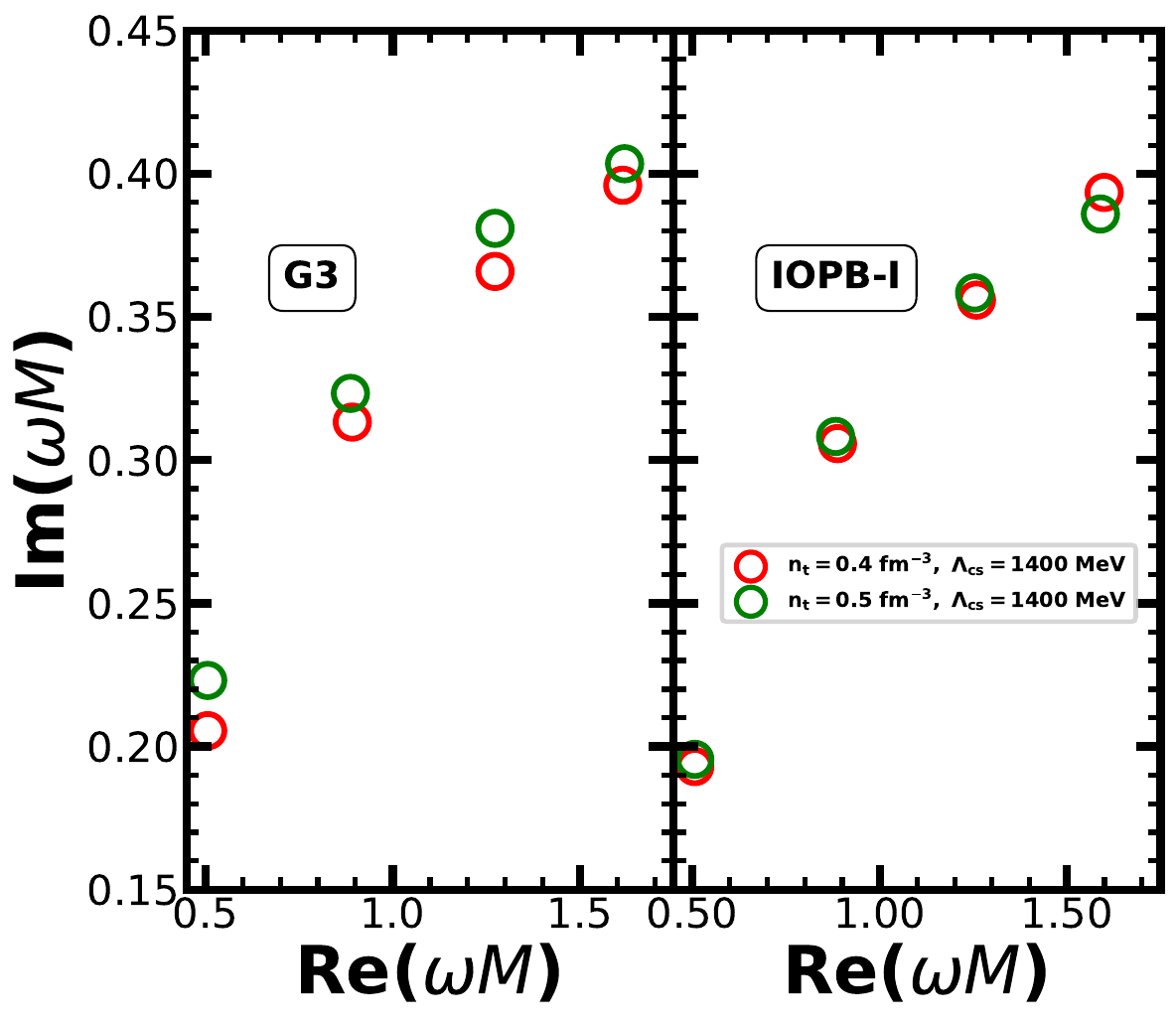}
        \caption{The dimensionless $\omega$ mode spectra ($\omega_1$, $\omega_2$, $\omega_3$, $\omega_4$) for G3 (left panel) and IOPB-I (right panel) parameter sets for quarkyonic star.}
        \label{fig: spectrum_transitiondensity}
    \end{figure}

Figure \ref{fig: spectrum_transitiondensity} presents the dimensionless  $\omega$-mode spectra for the same two RMF parameterizations (G3 and IOPB-I) as a function of the transition density $n_t$ i.e. the density at which quarkyonic degrees of freedom begin to contribute within the stellar core. The $n_t$ controls the quark content inside the star where the crossover physics become relevant. When we increase $n_t$ it delays the onset of quarkyonic behavior to higher densities. On the other hand, a smaller $n_t$ allows the quarkyonic sector to influence a larger fraction of the interior. Because  $\omega$-modes are predominantly spacetime oscillations that sense the integrated curvature profile. Thus the changes in the radial stratification of energy density and pressure due to $n_t$ manifest as systematic displacements of the complex eigenfrequencies. This behavior is evident in Fig.~\ref{fig: spectrum_transitiondensity}, where the overtone sequences again appear approximately uniformly spaced, which is consistent with a barrier-scattering interpretation of the QNMs. In particular, increasing $n_t$ lowers the real part of the eigenfrequency, indicating that postponing the quarkyonic onset tends to reduce the characteristic ringing frequency of the spacetime modes for the canonical configuration. The imaginary parts also adjust, reflecting changes in the efficiency with which the mode couples to outgoing gravitational radiation. Physically, this can be understood as a consequence of modifying the stellar compactness profile and the curvature potential that regulates partial trapping versus leakage of gravitational waves. As in Fig.~\ref{fig:spectrum_confinement}, the contrast between G3 and IOPB-I emphasizes that the same change in $n_t$ can produce quantitatively different spectral shifts depending on the underlying hadronic baseline.

\subsection{The analysis of $\omega_1$ and $\omega_2$ modes}

      \begin{figure}[h!]
        \centering
        \includegraphics[width=1\linewidth]{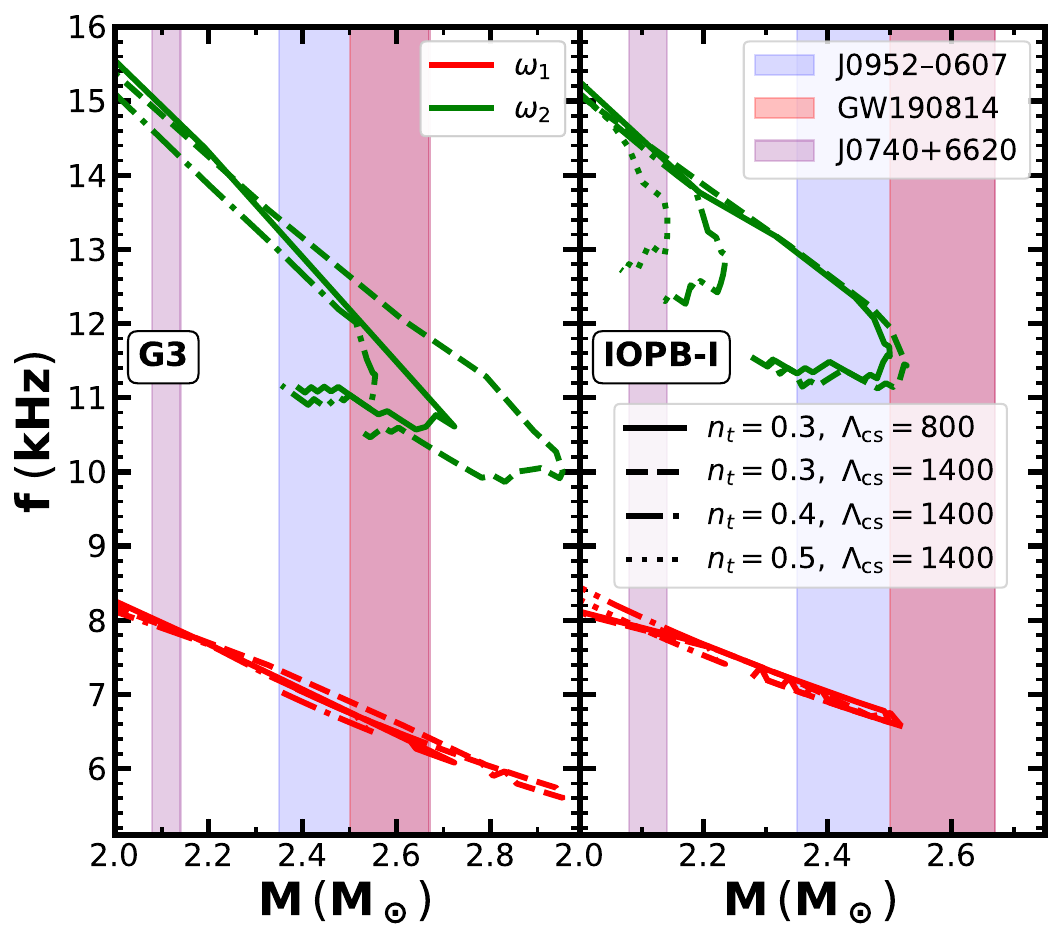}
        \caption{ The variation of fundamental ($\omega_1$, red color) and first overtone ($\omega_2$, green color) mode frequency with stellar mass for both G3 (left panel) and IOPB-I (right panel) parameter sets.}
        \label{fig:f_vs_M}
    \end{figure}
   Figure~\ref{fig:f_vs_M} correlates the  $\omega$ mode frequencies with the stellar mass along equilibrium sequences built from quarkyonic EOSs based on the G3 and IOPB-I RMF parameterizations. As spacetime-led modes, $\omega$ modes primarily track the star’s global compactness and the associated curvature potential. The overall trend is the configurations those are effectively more compact along a sequence exhibit higher oscillation frequencies. This behavior is consistent with the parameter-dependent trends summarized in Table~\ref{tab:dm_admixed_properties}. For a fixed RMF baseline, moving to higher transition densities $n_t$ is accompanied by an increase in the characteristic frequency $f$. For G3 at $\Lambda_{cs}=1400\,\mathrm{MeV}$, $f$ increases as ($5.612$, $6.467$,  $8.362)\,\mathrm{kHz}$ when $n_t$ increases as ($0.3$, $0.4$, $0.5)\,\mathrm{fm}^{-3}$. The corresponding $M_{\rm max}$ values are  ($2.95$, $2.56$, $1.58)\,M_\odot$. Conversely, increasing the confinement scale $\Lambda_{cs}$ at fixed $n_t$ tends to lower $f$. For G3 at $n_t=0.3\,\mathrm{fm}^{-3}$ changes from $f=6.021$ to $5.612\,\mathrm{kHz}$ when $\Lambda_{cs}$ is raised from $800$ to $1400\,\mathrm{MeV}$. These results reflects how the stiffening pattern of the quarkyonic sector modifies the curvature potential and shifts the mode spectrum. Comparing the two panels, the stiffer quarkyonic realizations (typically supporting larger radii and higher maximum masses in Table \ref{Tab: 1st}) systematically populate the lower-frequency part of Fig.~3, whereas softer realizations shift the spectrum to higher frequencies. Thus Fig.~3 and Table \ref{Tab: 1st} together demonstrate that the $\omega$ mode frequency provides a clean, EOS-sensitive diagnostic of how quarkyonic microphysics (through $n_t$ and $\Lambda_{cs}$) reshapes the bulk stellar structure.

\begin{figure}[h!]
        \centering
       \includegraphics[width=1\linewidth]{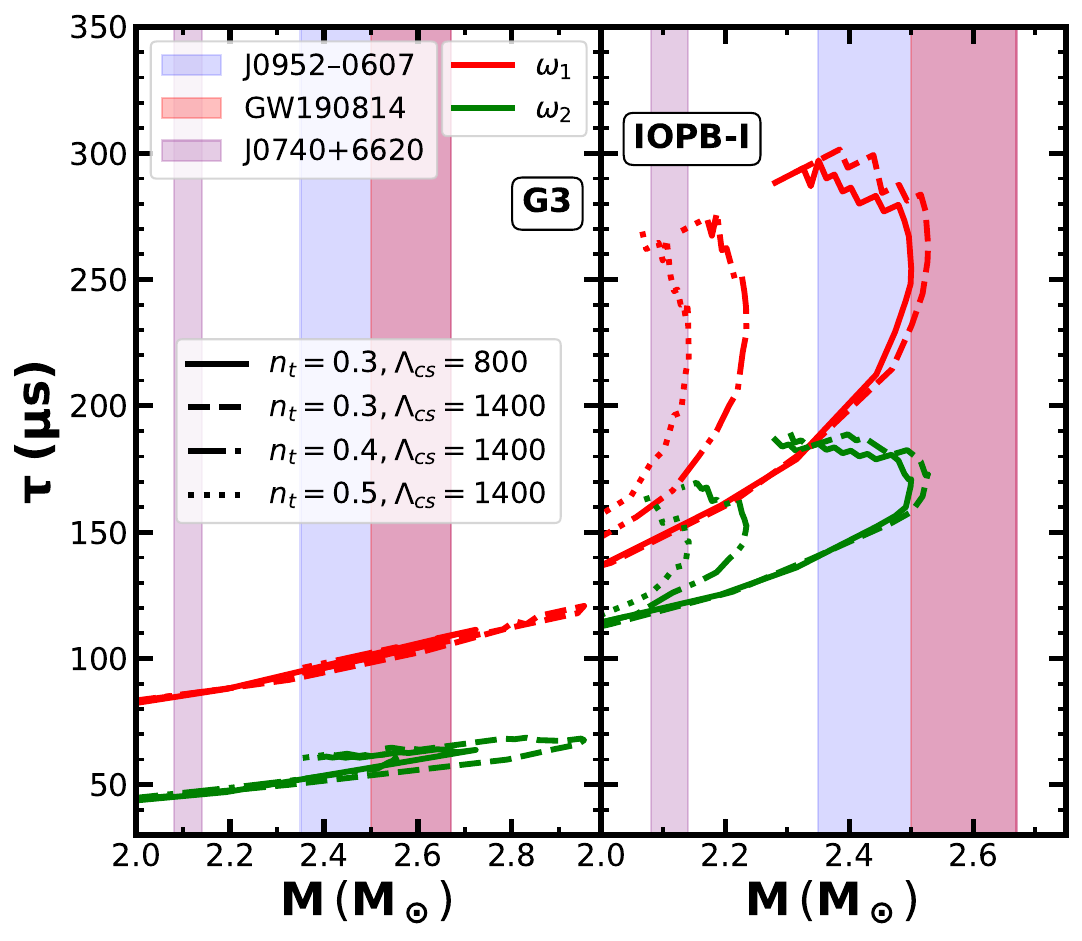}
        \caption{The variation of fundamental ($\omega_1$, red color) and first overtone ($\omega_2$, green color) damping time with stellar mass for both G3 (left panel) and IOPB-I (right panel) parameter sets.}
        \label{fig:tau_vs_M}
    \end{figure}

Figure~\ref{fig:tau_vs_M} shows the corresponding damping times $\tau$ of the $\omega$ modes as functions of stellar mass. It provides complementary information to Fig.~3 because $\tau$ measures the efficiency with which the spacetime oscillation radiates gravitational-waves. In general, more compact configurations radiate more efficiently and therefore have shorter damping times. The systematic parameter dependence of the physical quantities as shown in  Table~\ref{tab:dm_admixed_properties} supports this interpretation that for both RMF baselines, increasing $n_t$ produces a marked reduction in $\tau$. For G3 at $\Lambda_{cs}=1400\,\mathrm{MeV}$, $\tau$ decreases as $323.871$ , $273.848$, $192.101 \,\mu\mathrm{s}$ as $n_t$ increases from $0.3$ to $0.4$ and $0.5\,\mathrm{fm}^{-3}$ respectively. This indicates faster damping as the EOS becomes effectively softer and the curvature coupling strengthens. By contrast, increasing $\Lambda_{cs}$ at fixed $n_t$ increases $\tau$ (e.g., for IOPB-I at $n_t=0.3\,\mathrm{fm}^{-3}$, $\tau$ rises from $254.384$ to $264.916\,\mu\mathrm{s}$ when $\Lambda_{cs}$ is raised from $800$ to $1400\,\mathrm{MeV}$). This is consistent with stiffer quarkyonic realizations yielding longer-lived spacetime oscillations. In both Fig.~\ref{fig:f_vs_M} and Fig.~\ref{fig:tau_vs_M}, the fundamental $\omega_1$ branch and the first overtone $\omega_2$ branch remain clearly separated along the full mass range. At fixed $M$, the overtone has a higher frequency ($f_{w_2}>f_{w_1}$), and it typically damps (shorter $\tau$), reflecting the stronger radiative coupling of higher-order spacetime oscillations.

Recent multimessenger constraints provide complementary ``anchors'' for viable dense-matter equations of state. Precision radio timing has established the existence of neutron stars with gravitational masses at or above $2\,M_\odot$ (e.g., the $\sim 2\,M_\odot$ class of heavy pulsars) \cite{Demorest_2010,Antoniadis_2013,Cromartie_2020,Romani_2022}, while \textit{NICER} pulse-profile modelling constrains radii at the canonical mass scale ($M \simeq 1.4\,M_\odot$) to be $\mathcal{O}(12$--$13)\,\mathrm{km}$ \cite{Riley_2019,Miller_2019}. The gravitational-wave observations, notably GW170817 \cite{GW170817_binary_props}, further restrict the tidal deformability of $\sim 1.4\,M_\odot$ stars to the few-hundred level, disfavouring extremely stiff EOSs that would yield very large radii and correspondingly large $\Lambda_{1.4}$ \cite{PhysRevLett.119.161101,PhysRevLett.121.091102,PhysRevLett.121.161101,Capano2020}. In addition, the secondary component of GW190814, with a mass in the range $\sim 2.5$--$2.7\,M_\odot$, is often used as an empirical upper target for the maximum mass of nonrotating configurations, although its true nature (a massive neutron star versus a low-mass black hole) remains uncertain \cite{Abbott_2020}. Accordingly, an EOS capable of reaching into this mass window without violating radius and tidal-deformability constraints is especially compelling. In quarkyonic star models, the maximum masses listed in Table~1 span $M_{\max} \simeq 2.50$--$2.95\,M_\odot$ for representative choices of $(n_t,\Lambda_{cs})$ and for the two RMF baselines (G3 and IOPB-I). It  demonstrates  that quarkyonic stiffening at intermediate densities can naturally satisfy the robust $2\,M_\odot$ constraint and for certain parameter combinations, can even approach the GW190814 mass scale. The variations in $n_t$ and $\Lambda_{cs}$ can alter $M_{\max}$. This allows the quarkyonic EOS to remain compatible with the combined radius–tidal bound and support heavy mass neutron stars.
    
\begin{table}[h!]
	\centering
	\caption{Quarkyonic star properties such as maximum mass ($M_{max}$), maximum radius ($R_{max}$), compactness  ($C$), fundamental $\omega$ mode frequency ($f_{\omega_1}$), fundamental $\omega$ mode damping time ($\tau_{\omega_1}$) for different choices of transition density ($n_t$) and confinement scale ($\Lambda_{cs}$) for G3 and IOPB-I parameter sets.}
	\label{tab:dm_admixed_properties}
	\begin{tabular}{l c c c c c c c}
		\hline
		Model & $n_t$ & $\Lambda_{cs}$ & $M_{\rm max}$ & $R_{\rm max}$ & $C$ & $f_{\omega_1}$ & $\tau_{\omega_1}$ \\
		& (fm$^{-3}$) & (MeV) & ($M_\odot$) & (km) & & (kHz) & ($\mu$s) \\
		\hline
		G3 & 0.3 & 800 & 2.75 & 14.54 & 0.279 & 6.221 & 110.074 \\
		G3 & 0.3 & 1400 & 2.95 & 15.16 & 0.287 & 5.723 & 121.201 \\
		G3 & 0.4 & 1400 & 2.56 & 13.61 & 0.277 & 6.567 & 104.411 \\
		G3 & 0.5 & 1400 & 1.98 & 11.33 & 0.257 & 8.532 & 82.124 \\
		\hline
		IOPB-I & 0.3 & 800 & 2.50 & 13.65 & 0.270 & 6.734 & 287.704 \\
		IOPB-I & 0.3 & 1400 & 2.54 & 13.76 & 0.272 & 6.642 & 302.212 \\
		IOPB-I & 0.4 & 1400 & 2.24 & 12.30 & 0.268 & 7.623 & 279.701 \\
		IOPB-I & 0.5 & 1400 & 2.15 & 11.93 & 0.265 & 7.871 & 246.521 \\
		\hline
	\end{tabular}
    \label{Tab: 1st}
\end{table}
    
\begin{table}
\centering
\caption{Fitting results for frequency $f$ as a function of compactness $C$ using $f(C) = (1/R)[aC + b]$, where $R$ is the stellar radius. Results include data from G3 and IOPB-I equation of states.}
\begin{tabular}{l c c}
\hline
 & $\omega_1$ mode & $\omega_2$ mode \\
\hline
$a$ & $-3.86 \times 10^{2} \pm 2.15$ & $-8.97 \times 10^{2} \pm 1.09 \times 10^{1}$ \\
$b$ & $1.96 \times 10^{2} \pm 5.69 \times 10^{-1}$ & $3.97 \times 10^{2} \pm 2.89$ \\
$\chi^{2}$ & $2.28 \times 10^{2}$ & $2.28 \times 10^{2}$ \\
\hline
\end{tabular}
\label{tab:freq_fits}
\end{table}
 \begin{table}
	\centering
	\caption{Fitting results for damping time $\tau$ as a function of compactness $C$ using $\tau(C) = M/[aC^2 + bC + c]$, where $M$ is the stellar mass. Results include data from G3 and IOPB-I equation of states. }
	\begin{tabular}{l c c}
		\hline
		& $\omega_1$ mode & $\omega_2$ mode \\
		\hline
		$a$ & $7.70 \pm 2.22$ & $5.34 \pm 2.03$ \\
		$b$ & $-4.07 \pm 1.18$ & $-2.71 \pm 1.05$ \\
		$c$ & $0.549 \pm 0.157$ & $0.364 \pm 0.136$ \\
		$\chi^{2}$ & $222.42$ & $221.58$ \\
	
		\hline
	\end{tabular}
	\label{tab:tau_fits_corrected}
\end{table}

         \begin{figure}[h!]
         \centering
         \includegraphics[width=1\linewidth]{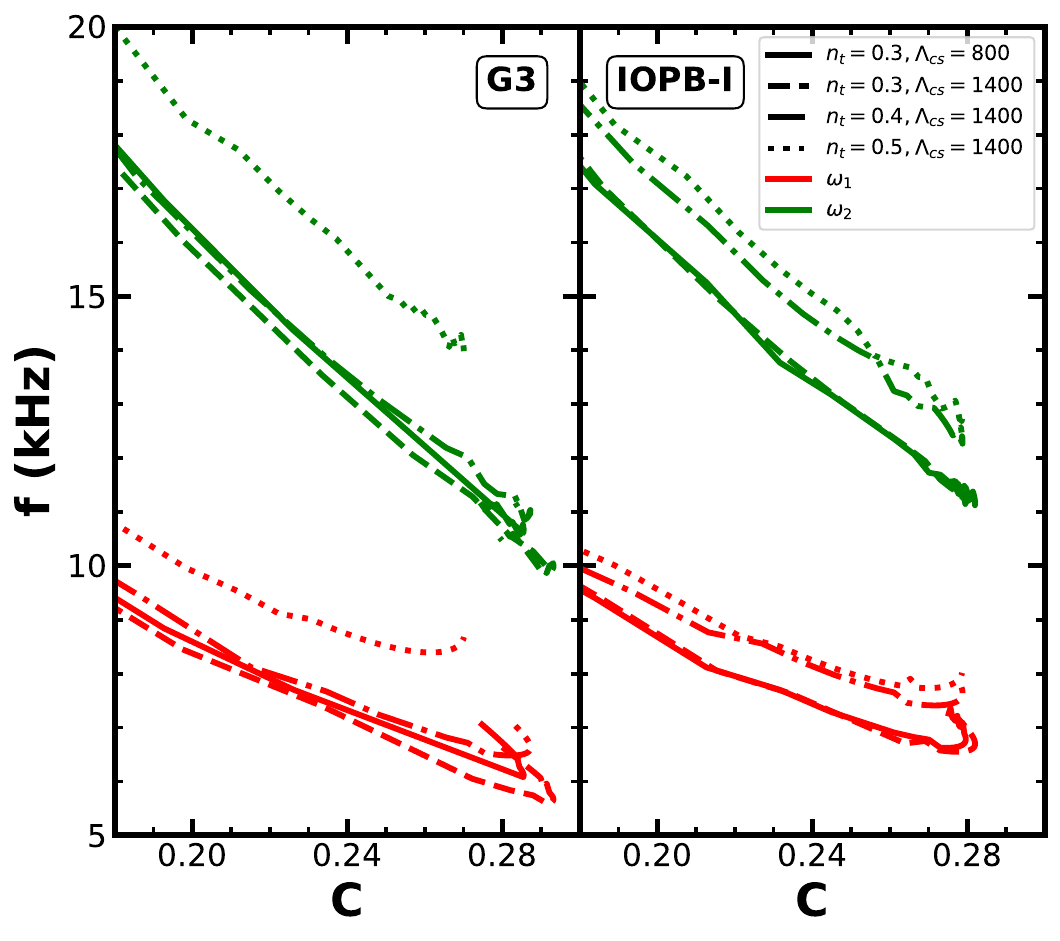}
         \caption{The variation of fundamental ($\omega_1$, red color) and first overtone ($\omega_2$, green color) mode frequency with compactness for both G3 (left panel) and IOPB-I (right panel) parameter sets.}
         \label{fig: f_vs_C}
     \end{figure}
    \begin{figure}[h!]
        \centering
        \includegraphics[width=1\linewidth]{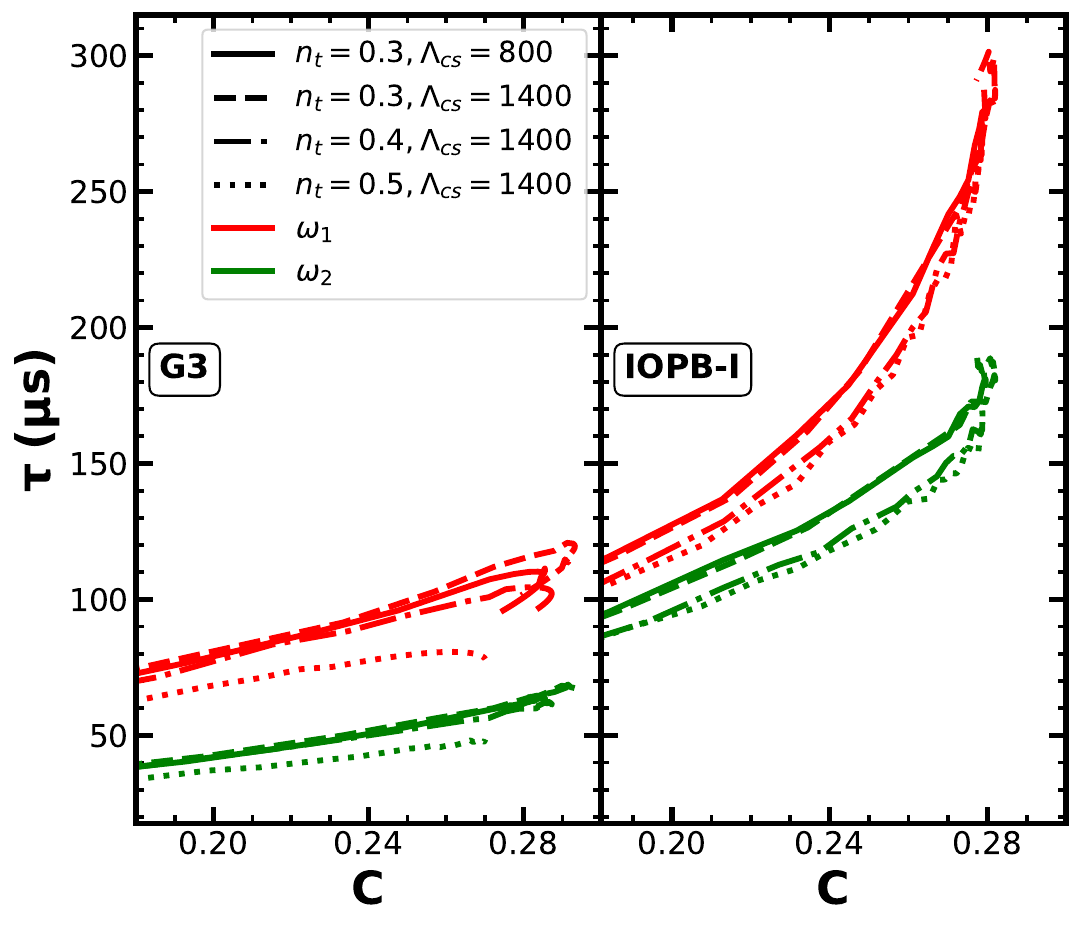}
        \caption{The variation of fundamental ($\omega_1$, red color)  and first overtone ($\omega_2$, green color) damping time with compactness for both G3 (left panel) and IOPB-I (right panel) parameter sets.}
        \label{fig: tau_vs_C}
    \end{figure}
 \begin{figure}
        \centering
        \includegraphics[width=1\linewidth]{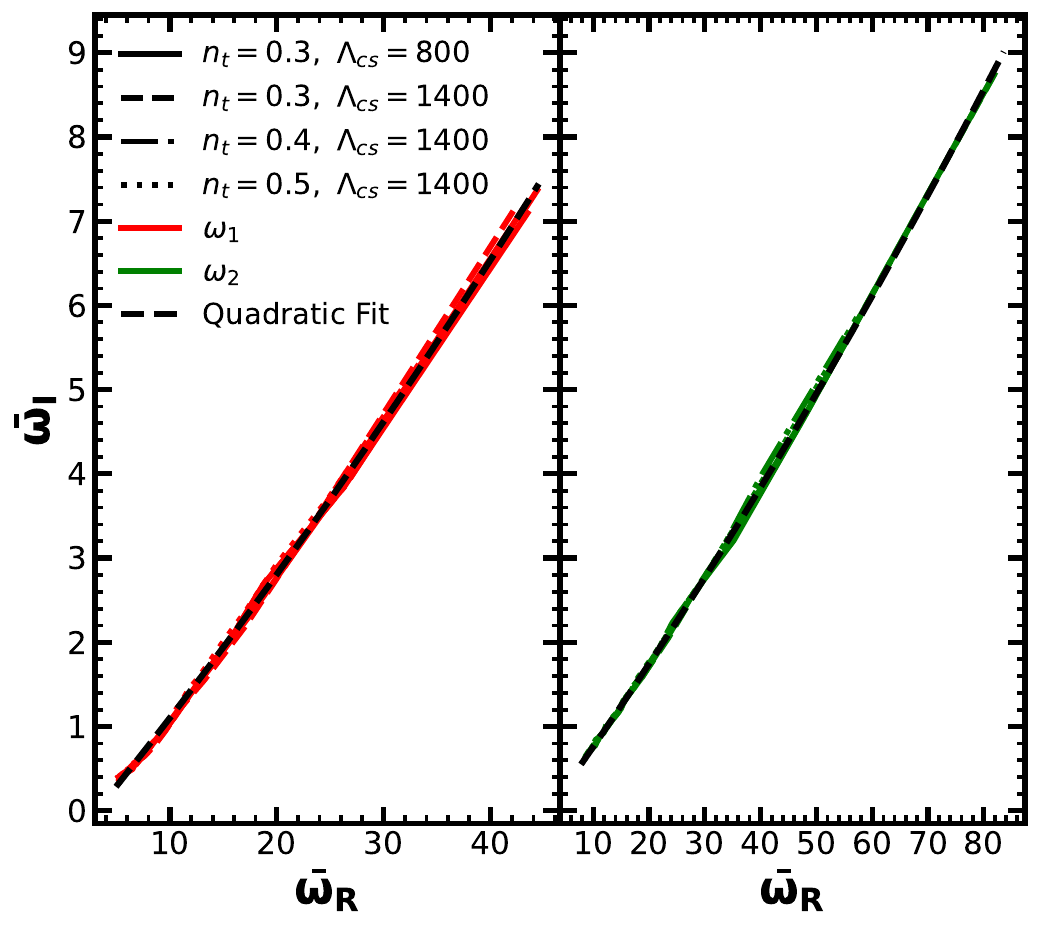}
        \caption{Universal relation for $\omega$ modes in the $(\bar{\omega}_{R},\bar{\omega}_{I})$ plane, where $\bar{\omega}_{R}$ and $\bar{\omega}_{I}$ are scaled (dimensionless) real and imaginary parts of the complex eigenfrequency defined using the central pressure $P_c$ (see Eq.~\eqref{eq:scaled_freqs_Pc}).}
        \label{fig:tilde_with_fits}
    \end{figure}

\subsection{The universal relations}

Figures~\ref{fig: f_vs_C} and \ref{fig: tau_vs_C}  highlight the dominant role of compactness $C\equiv M/R$ on the $\omega$ mode spectra for the G3 and IOPB-I parameter sets \cite{Debarati,Pheno_relations,Imprints_w_mode}. For both cases, the mode frequencies fundamental $\omega_1$ and first overtone $\omega_2$ decrease systematically with increasing $C$, while the corresponding damping times increase with $C$, reflecting how changes in the curvature potential and the exterior scattering barrier regulate both the oscillation scale and the leakage of gravitational radiation. To quantify these trends, we fit the fundamental and first-overtone branches with the empirical relations
\begin{equation}
 f(C)=\frac{1}{R}\left(a\,C+b\right),
\end{equation}
with best-fit dimensionless parameters $(a,b)$ given in Table~\ref{tab:freq_fits} and
\begin{equation}
 \tau(C)=\frac{M}{a\,C^{2}+b\,C+c},
\end{equation}
with parameters $(a,b,c)$ listed in Table~\ref{tab:tau_fits_corrected}. Here $M$ and $R$ are stellar mass and stellar radius, respectively. These compactness-based fits provide a convenient representation of the EOS dependence of the $\omega_1$ and $\omega_2$ branches and enable direct comparisons across different choices of quarkyonic parameters. Once $M$ and $R$ (hence $C$) are inferred from independent observations, the relations can be used to estimate $(f,\tau)$ for both the fundamental and first-overtone branches. The quality of the fits reported in Tables~\ref{tab:freq_fits} and \ref{tab:tau_fits_corrected} therefore quantifies the degree to which the $\omega$ mode spectra exhibit an approximately universal dependence on compactness for the present sets of quarkyonic EOS.

Figure \ref{fig:tilde_with_fits}  illustrates an empirical representation of the  $\omega$ mode spectrum in terms of  scaled (dimensionless) combinations of the real and imaginary parts of the complex eigenfrequency, expressed through the mode frequency $f$ and damping time $\tau$ \cite{ignacio_pressure_ur}. Following the definitions used in \cite{Pheno_relations}, we introduce the scaled quantities
\begin{align}
\bar{\omega}_{R} &\equiv \frac{2\pi f}{\sqrt{P_{c}}},
\\
\bar{\omega}_{I} &\equiv \frac{1/\tau}{\sqrt{P_{c}}},
\label{eq:scaled_freqs_Pc}
\end{align}
where $P_c$ is the central pressure of the stellar configuration and the prefactors/units are chosen such that $\bar{\omega}_{R}$ and $\bar{\omega}_{I}$ are dimensionless. 
The key point is that, after this rescaling by $\sqrt{P_c}$, results obtained for different EOSs (soft vs stiff) tend to collapse onto a 
common curve. This indicates that the central pressure encodes much of the relevant structural 
information controlling the spacetime-dominated oscillations.

To quantify this behavior in our calculations, we fit the scaled imaginary part $\bar{\omega}_{I}$ as a quadratic function of the scaled real part $\bar{\omega}_{R}$ defined as:
\begin{equation}
\bar{\omega}_{I} = A + B\,\bar{\omega}_{R} + C\,\bar{\omega}_{R}^{2},
\label{eq:global_quadratic_fit}
\end{equation}
with the best-fit coefficients $(A,B,C)$ for the fundamental ($\omega_1$) and first-overtone ($\omega_2$) branches as reported in 
Table~\ref{tab:wmode_global_fits}. Such empirical relations are potentially useful for neutron-star asteroseismology: if $f$ and $\tau$ 
are measured from a detected ringdown signal, Eq.~\eqref{eq:scaled_freqs_Pc} combined with the fit in Eq.~\eqref{eq:global_quadratic_fit} 
can be inverted to estimate the central pressure $P_c$ (or, equivalently, to place consistency constraints on candidate EOS models). 
In addition, the degree of scatter around the fitted curve provides a direct measure of how close the $\omega$ mode spectra is to being 
EOS-insensitive under the chosen scaling. The deviations (typically more pronounced for overtones) carry additional information about the 
detailed density profile and the sharpness of any phase transition in the core.\\

\begin{figure}
        \centering
        \includegraphics[width=1\linewidth]{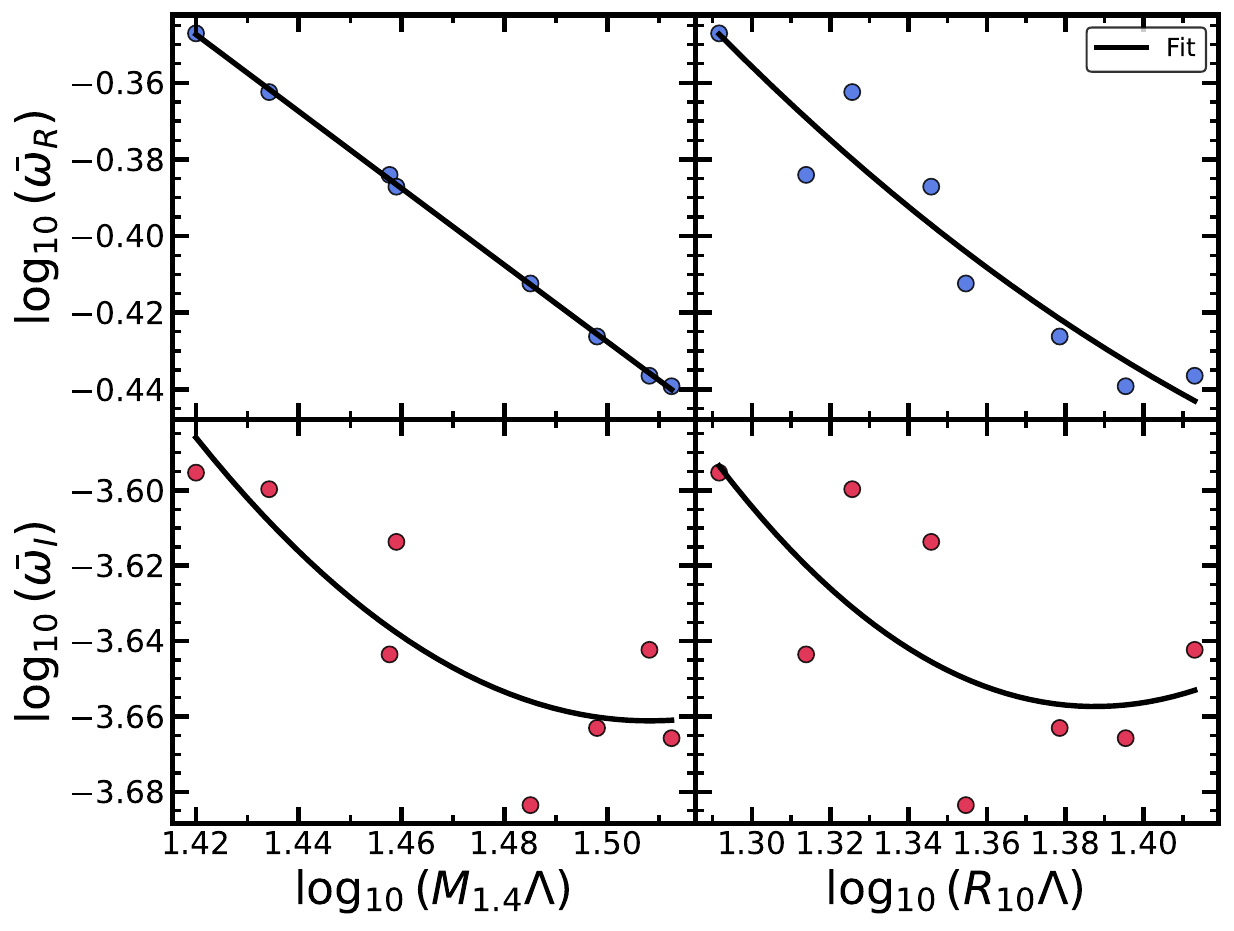}
        \caption{The plotted quantities are logarithms of the scaled real and imaginary parts of the mode frequency, $\log(\bar{\omega}_{R})$ and $\log(\bar{\omega}_{I})$, against (i) the combinations $M_{1.4}\Lambda$ (left panels) and (ii) the $R_{10}\Lambda$ (right panels), where $M_{1.4}\equiv M/(1.4M_{\odot})$, $R_{10}\equiv R/(10\,\mathrm{km})$, and $\Lambda$ is the dimensionless tidal deformability.}
        \label{fig:uv_tidal}
    \end{figure}

Figure~\ref{fig:uv_tidal} summarizes a set of empirical relationships proposed in the literature to connect the  $\omega$ mode spectra with tidal-deformability information. The figure indicates that the data collapse onto tight, nearly EOS-insensitive curves, motivating the use of simple fitting formulas. In particular, the best fits are given by
\begin{equation}
\log\left(\bar{\omega}_{R,I}\right)=a_{R,I}+b_{R,I}\,x+c_{R,I}\,x^{2},
\qquad x\equiv\log\left(M_{1.4}\Lambda\right),
\label{eq:UR_w0_x}
\end{equation}
with coefficients $(a_{R,I},b_{R,I},c_{R,I})$ listed in Table~\ref{tab:table1}, and
\begin{equation}
\log\left(\bar{\omega}_{R,I}\right)=\alpha_{R,I}+\beta_{R,I}\,y+\gamma_{R,I}\,\sqrt{y},
\qquad y\equiv\log\left(R_{10}\Lambda\right),
\label{eq:UR_w0_y}
\end{equation}
with coefficients $(\alpha_{R,I},\beta_{R,I},\gamma_{R,I})$ given in Table~\ref{tab:table2}. Although the solid lines in Fig. \ref{fig:uv_tidal} represent equations (\ref{eq:UR_w0_x}) and (\ref{eq:UR_w0_y}) with the coefficients given in Tables \ref{tab:table1} and  \ref{tab:table2} respectively, these lines are given just to guide the eyes. It is to be noted that there are some correlation of $\log_{10}(\bar{\omega}_{R})$ with $\log_{10}(M_{1.4}\Lambda)$ and $\log_{10}(R_{1.4}\Lambda)$ (upper panel of Fig. \ref{fig:uv_tidal}), but this correlation breaks completely for the imaginary component as shown in the lower part of the Fig. \ref{fig:uv_tidal} showing a larger error with the calculated results Tables \ref{tab:table1} and \ref{tab:table2}.

In the context of the present work, such universal relations are useful because they provide a compact bridge between spacetime QNM observables $(f,\tau)$ and tidal properties constrained during inspiral: if $\Lambda$ (or a narrow range of $\Lambda$) is inferred from a binary inspiral, then measuring a high-frequency ringdown feature consistent with an  $\omega$ mode could be cross-checked against Eqs.~(\ref{eq:UR_w0_x})--(\ref{eq:UR_w0_y}). Conversely, when combined with the results of quarkyonic-EOS for $\omega$ modes, deviations from these reference universal relation curves can indicate how additional microphysics (e.g., a crossover to quarkyonic matter controlled by $n_t$ and $\Lambda_{\rm cs}$) modifies the near-universality. This can potentially provides an additional diagnostic of the stellar interior; the numerical best-fit parameters used for these two universal relation are summarized in Tables~\ref{tab:table1} and \ref{tab:table2}.

 A meaningful interpretation of the universal relations obtained in this work requires a comparison with previous studies based on different compositions of dense matter. For purely hadronic equations of state, axial $\omega$-mode frequencies typically lie in the range $\sim 5$--$12~\mathrm{kHz}$, with damping times of order $\sim 10^{-4}$--$10^{-5}~\mathrm{s}$, depending on the stellar compactness \cite{Imprints_w_mode,Debarati}. It is well established that these quantities depend sensitively on the stiffness of the EOS. With softer composition, such as those including hyperons or meson condensates leads to higher frequencies (by $\sim 10$--$20\%$) and shorter damping times \cite{Debarati}. At the same time, approximate empirical relations linking mode properties to global stellar parameters (e.g., compactness $M/R$) can be constructed with such models \cite{Imprints_w_mode}.

More recently, investigations of hybrid stars with sharp hadron--quark phase transitions have demonstrated that the validity of such universal relations is not guaranteed. In particular, when the conversion between hadronic and quark matter is slow, extended branches of stable hybrid configurations emerge at central densities $\gtrsim (5$--$10)\,n_0$ (with $n_0 \simeq 0.16~\mathrm{fm}^{-3}$), and previously established universal relations for axial $\omega$ modes can be violated at the level of $\sim 10$--$30\%$ \cite{ignacio_pressure_ur}. This behavior reflects the presence of a discontinuous phase transition and the associated nontrivial dynamics at the interface. Improved universal relations that explicitly incorporate such slow-stable hybrid stars have also been proposed, yielding mass and radius estimates \cite{Ignacio_tidal_ur}.

In contrast, the quarkyonic framework considered in the present work is based on a crossover transition between hadronic and quark degrees of freedom rather than a sharp first-order transition. As a consequence, we find that the $\omega$-mode spectrum retains approximate universality, but with systematic and characteristic shifts in both frequency and damping time. These deviations arise from the gradual stiffening of the EOS due to the emergence of quark degrees of freedom, which can increase the effective speed of sound to values $C_s^2 \sim 0.5$--$0.8$ in the inner core, without introducing discontinuities in thermodynamic quantities \cite{tinaki_ur}. This distinguishes quarkyonic stars from both purely hadronic models and hybrid stars with sharp interfaces.

From a microscopic perspective, these results suggest that the behavior of universal relations can serve as a diagnostic of the nature of dense matter in neutron star interiors. While strong violations of universality (at the level of $\gtrsim 20\%$) may indicate a first-order phase transition with a sharp interface, the smoother deviations ($\lesssim 15\%$) observed in our study are consistent with a crossover scenario. Therefore, a combined analysis of $\omega$-mode frequencies and damping times together with universal relations, may provide a powerful tool to probe the onset and character of quark degrees of freedom in compact stars and to shed light on the possible existence of quarkyonic matter in their cores.   

    \begin{table}[h!]
    \centering
    \caption{Global quadratic fits for the scaled imaginary frequency
        $\bar{\omega}_I$ as a function of the scaled real frequency
        $\bar{\omega}_R$ for the $\omega_1$ and $\omega_2$ modes, including all EOSs and
        parameter sets. The fit function is
        $\bar{\omega}_I = A + B\,\bar{\omega}_R + C\,\bar{\omega}_R^{\,2}$.}
    \label{tab:wmode_global_fits}
    \renewcommand{\arraystretch}{1.2}
    \resizebox{\columnwidth}{!}{%
    \begin{tabular}{lcc}
        \hline\hline
        & $\omega_1$ mode & $\omega_2$ mode \\
        \hline
        $A$ 
        & $-4.68 \times 10^{-1} \pm 1.31\times 10^{-2}$ 
        & $-1.54\times 10^{-1} \pm 1.15 \times 10^{0}$ \\
        
        $B$ 
        & $1.50 \times 10^{-1} \pm 1.94 \times 10^{-3}$ 
        & $8.90 \times 10^{-2} \pm 9.49\times 10^{0}$ \\
        
        $C$ 
        & $6.17 \times 10^{-4} \pm 4.56 \times 10^{-5}$ 
        & $2.68\times 10^{-4} \pm 1.16\times 10^{0}$ \\
        
        $\chi^{2}$ 
        & $1.02$ 
        & $2.06\times 10^{-1}$\\
		
        \hline\hline
    \end{tabular}%
    }
\end{table}

    \begin{table}[htbp]
	\centering
	\caption{Fitting results for $\log_{10}(\bar{\omega}_{R,I})$ as a function of $\log_{10}(M_{1.4}\Lambda)$ as defined in equation (\ref{eq:UR_w0_x})  with coefficients $(a_{R,I},b_{R,I},c_{R,I}).$  }
	\label{tab:table1}
	\begin{tabular}{lcc}
		\toprule
		Parameter & Value ($\pm$ error) & Error (\%) \\
		\midrule
		$a_R$ & $1.197 \pm 0.850$ & 71.00 \\
		$b_R$ & $-1.166 \pm 1.158$ & 99.27 \\
		$c_R$ & $0.056 \pm 0.394$ & 707.78 \\
		$a_I$ & $18.349 \pm 18.633$ & 101.54 \\
		$b_I$ & $-29.189 \pm 25.393$ & 87.00 \\
		$c_I$ & $9.677 \pm 8.648$ & 89.37 \\
		\bottomrule
	\end{tabular}
\end{table}

\begin{table}[htbp]
	\centering
	\caption{Fitting results for $\log_{10}(\bar{\omega}_{R,I})$ as a function of $\log_{10}(R_{1.4}\Lambda)$ as defined in equation (\ref{eq:UR_w0_y}) with coefficients $(\alpha_{R,I},\beta_{R,I},\gamma_{R,I})$.}
	\label{tab:table2}
	\begin{tabular}{lcc}
		\toprule
		Parameter & Value ($\pm$ error) & Error (\%) \\
		\midrule
		$\alpha_R$ & $14.894 \pm 23.715$ & 159.22 \\
		$\beta_R$ & $9.739 \pm 17.535$ & 180.04 \\
		$\gamma_R$ & $-24.480 \pm 40.787$ & 166.61 \\
		$\alpha_I$ & $47.937 \pm 52.313$ & 109.13 \\
		$\beta_I$ & $37.184 \pm 38.680$ & 104.02 \\
		$\gamma_I$ & $-87.602 \pm 89.974$ & 102.71 \\
		\bottomrule
	\end{tabular}
\end{table}

\section{Summary and Conclusions}
\label{Concl.}
  
This work has investigated spacetime-led quasinormal modes of relativistic compact stars. The stellar interior is modeled using quarkyonic matter constructed within relativistic mean-field theory. We adopt the G3 and IOPB-I parameterizations as hadronic baselines \cite{Kumar_2017,Kumar_2018} EOS. The quarkyonic crossover is controlled by two high-density parameters---the transition density $n_t$ and the confinement scale $\Lambda_{\rm cs}$. These are varied to generate the EOS families compatible with standard mass--radius constraints \cite{Demorest_2010,Antoniadis_2013,Riley_2019,Miller_2019,PhysRevLett.119.161101}. For each EOS realization, the complex eigenfrequencies are computed in full general relativity and reported through the mode frequency and damping time.\\Across the quarkyonic-RMF families considered here, the $\omega$ mode spectrum exhibits a coherent, EOS-dependent shift under variations in $(n_t,\Lambda_{\rm cs})$. In line with established results for spacetime modes \cite{Anderson_1998,Leins_1993}, the dominant control parameters are the stellar compactness and the associated effective curvature potential. These govern gravitational-wave scattering in the exterior region near the stellar surface \cite{Leins_1993,Benhar_2004}. Modifying $n_t$ and $\Lambda_{\rm cs}$ changes the density profile and compactness in a correlated way and therefore displaces the full QNM ladder in the complex-frequency plane. The fundamental and first-overtone branches remain clearly separated. The overtone shows higher oscillation frequencies and typically shorter damping times. This is consistent with the expected ordering of higher-curvature oscillations \cite{Anderson_1998,Leins_1993}.\\To support data-driven applications, compact empirical representations were provided for the dependence of $f$ and $\tau$ on the compactness $C\equiv M/R$ for both the fundamental ($\omega_1$) and first-overtone ($\omega_2$)  modes (Tables~\ref{tab:freq_fits} and \ref{tab:tau_fits_corrected}). In addition, scaled-frequency representations and a global quadratic relation between $\bar{\omega}_I$ and $\bar{\omega}_R$ were constructed (Table~\ref{tab:wmode_global_fits}) \cite{Leins_1993}. Such relations offer a practical mapping between potentially observable high-frequency ringdown features and bulk stellar properties (e.g., $M$, $R$, and tidal deformability). We have studied various universal relations, and our results reveal distinctive features of quarkyonic stars. From an observational perspective,  $\omega$ modes occupy a high-frequency band (typically $\sim 5$--$20\,\mathrm{kHz}$) and are strongly damped (typically $\tau\sim 10^{-4}\,\mathrm{s}$), making detectability challenging for current interferometers \cite{Anderson_1998,Benhar_2004}. Nevertheless, proposed excitation channels include neutron-star collapse to a black hole and strong dynamical phases in compact-object evolution, where rapid spacetime ringing can be triggered \cite{Benhar_2004}. Improved high-frequency sensitivity in future detector generations may therefore enable spacetime-mode astroseismology as a complementary EOS diagnostic. This approach can be further strengthened when combined with independent constraints from inspiral measurements \cite{PhysRevLett.119.161101,PhysRevLett.121.091102,Capano2020}. \\ 

Since the internal structure of neutron star is highly complex and cannot be fully incorporated within the scope of the present study, several natural extensions to this work arise. These include incorporating rotation and magnetic fields to quantify their impact on the spectrum, as well as assessing the quality of the empirical fits \cite{Anderson_1998,Leins_1993}. They also involve expanding the analysis to additional EOS families and alternative high-density scenarios to test the robustness of the inferred trends \cite{Masuda2013,McLerran_2019}. Another important direction is to integrate the present quarkyonic-mode predictions with numerical merger and collapse simulations in order to estimate realistic excitation amplitudes. This synergy would pave the way for performing Bayesian inference of 
($n_t$, $\Lambda_{cs}$) using future high-frequency detections \cite{PhysRevLett.119.161101,Benhar_2004}.

\appendix
\section{Phase amplitude method}
\label{appendix}

Equation (\ref{zerilli_eq}) bears a formal resemblance to the time-independent Schrödinger equation; obtaining accurate solutions for quasinormal mode frequencies is a highly non-trivial numerical undertaking. The primary challenge lies in the precise implementation of the physical boundary condition requiring purely outgoing gravitational waves at spatial infinity. This requirement must be translated into a numerical computation in two problematic steps: first, one must approximate ``infinity" by a finite but large radial coordinate, and second—constituting the major numerical difficulty—one must clearly separate two linearly independent solutions whose asymptotic behavior is exponentially growing and decaying, respectively. This numerical instability, where tiny errors in the decaying solution can be overwhelmed by contamination from the growing solution, has spawned the development of specialized techniques such as Leaver’s continued-fraction method and various WKB/phase-integral approximations \cite{leaver_QNM_techniques_1985,Kokkotas_living_review}.

In the present work, we employ the phase-amplitude method, as formulated by Anderson et al.~\cite{Anderson_inverse_cowling}. This method is originally developed and demonstrated in the calculation of black hole quasinormal modes. Its application to the neutron star problem is well-motivated, as it directly addresses the core numerical difficulty. The method's advantage is highlighted in comparative studies where traditional phase-integral approaches, such as the one derived by Fr\"oman et al.\cite{Fromen}, were found to yield reliable frequencies only for the very lowest-order modes. In contrast, the phase-amplitude method is shown to generate highly accurate normal-mode frequencies across a wide spectrum. It achieves this robustness by reformulating the problem in terms of a phase function and its derivative, which remain well-behaved numerically even where the wavefunction itself is not. This stability allows for a more precise determination of the complex eigenfrequencies that satisfy the outgoing-wave boundary condition. The following section provides a detailed description of the phase-amplitude formalism and its specific implementation for calculating the quasinormal modes of neutron stars. 
 The phase-amplitude method, as introduced by Anderson et al.~\cite{Anderson_inverse_cowling}, provides a robust framework for overcoming the numerical challenges inherent in solving the Zerilli equation for quasinormal modes. The method begins with a transformation of the dependent variable designed to simplify the asymptotic behavior of the solutions. Specifically, one defines a new function \(\Psi\) related to the Zerilli function \(Z\) by:
\begin{equation}
Z = \left(1 - \frac{2M}{r}\right)^{-1/2} \Psi.
\end{equation}
This transformation removes the first derivative term from the resulting wave equation, yielding a Schrödinger-like form:
\begin{equation}
\left( \frac{d^2}{dr^2} + U(r) \right) \Psi = 0,
\end{equation}
where the effective potential \(U(r)\) incorporates the original Zerilli potential \(V_Z(r)\) along with additional terms arising from the coordinate transformation. The key insight is to express the two linearly independent solutions \(\Psi^\pm\) in a phase-amplitude form:
\begin{equation}
\Psi^\pm = q^{-1/2} \exp\left[ \pm i \int^r q(\hat{r}) \, d\hat{r} \right],
\end{equation}
where the function \(q(r)\) satisfies the nonlinear differential equation\cite{Anderson_inverse_cowling}:
\begin{equation}
\frac{1}{2q} \frac{d^2 q}{dr^2} - \frac{3}{4q^2} \left( \frac{dq}{dr} \right)^2 + q^2 - U = 0.
\end{equation}
Although this equation is nonlinear and appears more complex, it possesses significant numerical advantages. While the original wavefunction \(\Psi\) oscillates rapidly, especially for high frequencies, the function \(q(r)\) is typically slowly varying. This slow variation makes \(q(r)\) amenable to stable numerical integration. Initial conditions for \(q\) can be generated using the WKB approximation \(q \approx \sqrt{U}\) at a large distance from the star, where \(U\) varies slowly.

The phenomenon of Stokes lines plays a crucial role in understanding the behavior of the solutions and must be accounted before numerical integration \cite{Anderson_inverse_cowling}. Stokes lines are curves in the complex plane emanating from turning points (zeros of \(U\)) where the asymptotic behavior of the WKB solutions changes discontinuously. When crossing a Stokes line, the coefficient of the subdominant exponential solution can change abruptly—a phenomenon known as the Stokes phenomenon \cite{Anderson_inverse_cowling}. This means that a linear combination of \(\Psi^+\) and \(\Psi^-\) that represents the physical solution on one side of a Stokes line may not be valid on the other side. 

To handle this properly, one must identify the appropriate anti-Stokes lines in the complex \(r\)-plane. Anti-Stokes lines are curves along which the phase integral \(\int q \, dr\) is purely real, ensuring that the solutions remain oscillatory and bounded. For quasinormal modes with complex frequency \(\omega\), the optimal integration path is a straight line with slope given by \(\tan \theta = -\text{Im} \, \omega / \text{Re} \, \omega\), which aligns with an anti-Stokes line. This path choice is essential for suppressing the exponential divergence of the outgoing wave solution and obtaining numerically stable results. The derivative along this path is computed using:
\begin{equation}
\frac{dq}{dr} = e^{-i\theta} \frac{dq}{d\rho},
\end{equation}
where \(\rho\) is the real distance along the integration path. The phase-amplitude method inherently accounts for the Stokes phenomenon by working with the slowly varying \(q\)-function, which remains smooth across Stokes lines, thereby avoiding the discontinuities that come from WKB approaches.

The numerical solution proceeds by integrating the \(q\)-equation from a large complex \(r\) (where the WKB initial conditions are valid) inward along the chosen anti-Stokes line to the stellar surface \(r = R\). At the surface, the exterior solution must match the interior solution. The matching condition leads to an expression for the amplitude ratio of incoming to outgoing waves. This ratio can be expressed as \cite{Anderson_inverse_cowling}:
\begin{equation}
\label{eq:A6}
\mathcal{R}(\omega) = \frac{ 
	\mathcal{Z}_2 \left[ \left(1 - \frac{2M}{R} \right) \left( i q + \frac{1}{2q} \frac{dq}{dr} \right) + \frac{M}{R^2} \right] + \frac{d\mathcal{Z}_2}{dr}
}{ 
	\mathcal{Z}_2 \left[ \left(1 - \frac{2M}{R} \right) \left( i q - \frac{1}{2q} \frac{dq}{dr} \right) - \frac{M}{R^2} \right] - \frac{d\mathcal{Z}_2}{dr}
},
\end{equation}
where \(\mathcal{Z}_2\) represents the interior solution evaluated at the stellar surface, and all quantities are computed at \(r = R\). The quasinormal modes are precisely those complex frequencies \(\omega_n\) for which this ratio vanishes, \(\mathcal{R}(\omega_n) = 0\), indicating no incoming radiation. This condition is solved iteratively, and the use of the complex path along anti-Stokes lines ensures that the exponentially growing component is controlled, allowing for accurate determination of the mode frequencies. This approach has proven highly effective for both black hole and neutron star perturbations, providing reliable results even for highly damped modes where traditional WKB methods fail. 
\bibliography{wmode}
\bibliographystyle{apsrev4-1}
\end{document}